\documentclass{aa}  

\usepackage{graphicx}
\usepackage{lscape}
\usepackage{longtable}
\usepackage{txfonts}
\begin{document}

\title{Origin of central abundances in the hot intra-cluster medium}
\subtitle{I. Individual and average abundance ratios from \textit{XMM-Newton} EPIC}

\author{F. Mernier\inst{\ref{SRON},\ref{Leiden}} \and J. de Plaa\inst{\ref{SRON}} \and C. Pinto\inst{\ref{Cambridge}}  \and J. S. Kaastra\inst{\ref{SRON},\ref{Leiden}} \and P. Kosec\inst{\ref{Cambridge}} \and Y.-Y. Zhang\inst{\ref{Bonn}} \and J. Mao\inst{\ref{SRON},\ref{Leiden}} \and N. Werner\inst{\ref{Stanford1},\ref{Stanford2}}}

\institute{SRON Netherlands Institute for Space Research, Sorbonnelaan 2, 3584 CA Utrecht, The Netherlands \\ \email{F.Mernier@sron.nl}\label{SRON} \and Leiden Observatory, Leiden University, P.O. Box 9513, 2300 RA Leiden,The Netherlands\label{Leiden} \and Institute of Astronomy, Madingley Road, CB3 0HA Cambridge, United Kingdom\label{Cambridge} \and Argelander-Institut f\"{u}r Astronomie, Auf dem H\"{u}gel 71, D-53121 Bonn, Germany\label{Bonn} \and Kavli Institute for Particle Astrophysics and Cosmology, Stanford University, 452 Lomita Mall, Stanford, CA 94305, USA\label{Stanford1} \and Department of Physics, Stanford University, 382 Via Pueblo Mall, Stanford, CA 94305-4060, USA\label{Stanford2}}

\date{Received 24 November 2015 / Accepted 3 June 2015}

\abstract{The hot intra-cluster medium (ICM) is rich in metals, which are synthesised by supernovae (SNe) explosions and accumulate over time into the deep gravitational potential well of clusters of galaxies. Since most of the elements visible in X-rays are formed by type Ia (SNIa) and/or core-collapse (SNcc) supernovae, measuring their abundances gives us direct information on the nucleosynthesis products of billions of SNe since the epoch of the star formation peak ($z \sim $ 2--3). In this study, we use the EPIC and RGS instruments on board \textit{XMM-Newton} to measure the abundances of nine elements (O, Ne, Mg, Si, S, Ar, Ca, Fe, and Ni) from a sample of 44 nearby cool-core galaxy clusters, groups, and elliptical galaxies. We find that the Fe abundance shows a large scatter ($\sim$20--40\%) over the sample, within $0.2r_{500}$ and especially $0.05r_{500}$. Unlike the absolute Fe abundance, the abundance ratios (X/Fe) are  uniform over the considered temperature range ($\sim$0.6--8 keV) and with a limited scatter. In addition to an unprecedented treatment of systematic uncertainties, we provide the most accurate abundance ratios measured so far in the ICM, including Cr/Fe and Mn/Fe which we firmly detected (>4$\sigma$ with MOS and pn independently). We find that Cr/Fe, Mn/Fe, and Ni/Fe differ significantly from the proto-solar values. However, the large uncertainties in the proto-solar abundances prevent us from making a robust comparison between the local and the intra-cluster chemical enrichments. We also note that, interestingly, and despite the large net exposure time ($\sim$4.5 Ms) of our dataset, no line emission feature is seen around $\sim$3.5 keV.}


\maketitle

\titlerunning{Origin of central abundances in the hot intra-cluster medium I}
\authorrunning{F. Mernier et al.}

\section{Introduction}\label{sect:intro}

About 80--90\% of the baryonic matter in the Universe is in the form of a hot and diffuse intergalactic gas, which has mostly a very low density and, therefore, is hard to observe. However, in the largest gravitationally bound regions of the Universe, which are clusters of galaxies, the density and temperature of this hot gas, or intra-cluster medium (ICM), becomes high enough for it to glow in X-rays. This ICM, which has been extensively studied by X-ray observatories over the past decades \citep[for a review, see][]{2010A&ARv..18..127B}, is particularly rich in metals \citep[e.g.][]{1976MNRAS.175P..29M,1996ApJ...466..686M}. Since the baryonic content of the Universe just after the Big Bang  consists exclusively of hydrogen and helium (and traces of lithium), these heavy elements -- typically from oxygen to nickel -- must have been synthesised by stars and supernovae (SNe) in the galaxy members and then ejected into the ICM \citep[for a review, see][]{2008SSRv..134..337W,2013AN....334..416D}.

Although the general picture of this chemical enrichment is now well established, many aspects are still poorly understood. In addition to the question of the transport mechanisms that drive the enrichment, a major uncertainty resides in the metal yields produced by type Ia (SNIa) and core-collapse (SNcc) supernovae. In fact, the nature of the SNIa progenitors and the SNIa explosion mechanisms are still under debate, while the global nucleosynthesis of SNcc highly depends on the initial mass function (IMF) and the initial metallicity of the considered stellar population. Moreover, in addition to SNe, AGB stars can also play a role in releasing lighter metals (e.g. nitrogen) or even heavy metals (via the s-process). Taken together, these unsolved questions lead to large uncertainties in predicting the global abundance ratios that are finally released by the SNe and AGB stars into the ICM.

In contrast to the remaining uncertainties in the theoretical yields from the SNe/AGB models, the current generation of X-ray observatories  measures the chemical abundances in the ICM with  remarkable accuracy since most transitions of H- and He-like elements from $Z$=7 to $Z$=28 fall within 0.2--12 keV. Thanks to the large effective area and the good spectral resolution of its European Photon Imaging Camera \citep[EPIC,][]{2001A&A...365L..18S,2001A&A...365L..27T} and Reflection Grating Spectrometer \citep[RGS,][]{2001A&A...365L...7D}, \textit{XMM-Newton} is particularly suitable for measuring abundances of elements like oxygen (O), neon (Ne), magnesium (Mg), silicon (S), sulfur (S), argon (Ar), calcium (Ca), iron (Fe), and nickel (Ni), especially in cool-core objects\footnote{A cluster, or group, is defined as ``cool-core'' when the ICM in its core is sufficiently dense that its cooling time, typically of the order of $\sim \sqrt{T_X}/n_e$, is shorter than the Hubble time.} which exhibit a high surface brightness in X-ray and where the most prominent K-shell emission lines of these elements are clearly detected. Consequently, the accuracy of these measurements can, in principle, bring new constraints on the SNe (and AGB) models, and can lead to a deeper understanding of the chemical enrichment processes beyond galactic scales.

Several authors have reported such analyses by measuring the abundances in the ICM of nearby clusters and groups. For instance, \citet{2007A&A...465..345D} has compiled a sample of 22 cool-core clusters and found that the standard SNIa models fail to reproduce the Ar/Ca and Ca/Fe abundance ratios. They also show that the number of SNIa over the total number of SNe highly depends on the considered models. \citet{2009A&A...508..565D} have shown that Si/Fe abundance ratios are remarkably uniform over a sample of 26 cool-core clusters observed with \textit{XMM-Newton}, arguing for a similar enrichment process within cluster cores. However they suggest that systematic uncertainties are too large to precisely estimate the relative contribution of SNIa and SNcc. Finally, many abundance studies have also been performed on individual objects   \citep[e.g.][]{2006A&A...449..475W,2006A&A...452..397D,2007ApJ...667L..41S,2009A&A...493..409S,2015A&A...575A..37M}. From these studies, and considering the actual instrumental performances of current X-ray observatories, it appears that higher quality data (i.e. with longer exposure time) spread over larger samples are needed to clarify the actual picture of the precise origin of metals in the ICM.

In this work (hereafter Paper I), we use new and archival \textit{XMM-Newton} EPIC observations to measure the chemical abundances of nine elements (O, Ne, Mg, Si, S, Ar, Ca, Fe, and Ni) in the core of a sample of nearby cool-core galaxy clusters, groups, and elliptical galaxies. These EPIC observations are then combined with the RGS abundance measurements adapted from de Plaa et al. (to be submitted) in order to derive average X/Fe abundance ratios representative of the nearby ICM. Taking into account as many systematic uncertainties as possible, we discuss the robustness of these measurements and compare them to the proto-solar abundances. In a second paper \citep[][hereafter Paper II]{2016arXiv160803888M}, we  discuss in detail the astrophysical implications of our results, and compare our average abundance pattern presented here with predictions from theoretical SNIa and SNcc yield models.

This paper is structured as follows.
In Sect. \ref{sect:data_reduction}, we present the sample and the data reduction pipeline. In Sect. \ref{sect:spectral_analysis} we describe the spectral analysis procedure applied to our EPIC observations. Our results are presented in Sect. \ref{sect:results}, briefly discussed in Sect. \ref{sect:discussion}, and summarised in Sect. \ref{sect:conclusion}.
Throughout this paper we assume cosmological parameters of $H_0 = 70$ km s$^{-1}$ Mpc$^{-1}$, $\Omega_m = 0.3,$ and $\Omega_\Lambda = 0.7$. All the error bars are given at a $68\%$ confidence level.


\section{Observations and data preparation}\label{sect:data_reduction}


Our sample consists of the CHEERS\footnote{CHEmical Enrichment Rgs Sample} catalogue (de Plaa et al., to be submitted), and is detailed in Table \ref{table:observations} \citep[see also][and de Plaa et al., to be submitted]{2015A&A...575A..38P}. It includes 44 nearby ($z < 0.1$) cool-core clusters, groups, and elliptical galaxies for which the \ion{O}{viii} 1s--2p line at 19 $\AA$  is detected by the RGS instrument with >5$\sigma$. More information on the intrinsic properties of these objects (e.g. fluxes) can be found in various available cluster catalogues, such as the HIGFLUGCS \citep{2002ApJ...567..716R}, the REFLEX \citep{2004A&A...425..367B}, and the ACCEPT \citep{2009ApJS..182...12C} catalogues. In our sample, recent \textit{XMM-Newton} observations (AO-12, PI: de Plaa) have been combined with archival data. We only select the pointings for which the combined EPIC observations (MOS\,1, MOS\,2, and pn)  gather at least 15 ks of net exposure time. The observations that suffer from high soft flare events or calibration problems are also excluded.

\begin{table*}[!]
\begin{centering}
\setlength{\tabcolsep}{4.5pt}
\caption{\textit{XMM-Newton}/EPIC observations used in this paper \citep[see][for details on RGS observations]{2015A&A...575A..38P}. The new observations from the CHEERS proposal are shown in boldface.} 
\label{table:observations}
\begin{tabular}{l c c c c c c c}        
\hline \hline
Source & ObsID & \multicolumn{3}{c}{Net exposure time$^{(a)}$} & $z$$^{(b)}$ & $r_{500}$$^{(c)}$ & Type$^{(d)}$\\
 &  & MOS\,1 & MOS\,2 & pn &  &  &  \\
 &  & (ks) & (ks) & (ks) &  & (Mpc) &  \\

\hline 

2A 0335+096 &   0109870101 0147800201  & 85.7 & 86.6 & 81.4 & 0.0349 & 1.05 &  h \\ 

A\,85 &   \textbf{0723802101/2201}  & 191.1 & 193.6 & 158.2 & 0.0556 & 1.21 & h \\ 

A\,133 &   0144310101 \textbf{0723801301/2001}  & 134.3 & 137.4 & 95.8 & 0.0569  & 0.94 & h \\ 

A\,189 &   0109860101 & 35.7 & 36.8 & 33.3 & 0.0318 & 0.50 & c \\ 

A\,262 &   0109980101 0504780201 & 53.4 & 54.7 & 43.2 & 0.0161 &  0.74 & h \\ 

A\,496 &   0135120201 0506260301/0401 & 131.4 & 138.2 & 103.3 & 0.0328 & 1.00 & h \\ 

A\,1795 &   0097820101 & 38.1 & 37.0 & 25.4 & 0.0616 & 1.22 & h \\

A\,1991 &   0145020101 & 29.2 & 29.3 & 20.5 & 0.0587 & 0.82 & h \\

A\,2029 &   0111270201 0551780201/0301/0401/0501 & 147.5 & 154.4 & 107.7 & 0.0767 & 1.33 & h\\

A\,2052 &   0109920101 0401520501/0801 & 93.0 & 92.7 & 53.7 & 0.0348 & 0.95 & h \\
  &   0401520901/1101/1201/1601 &  &  &  &  &  \\

A\,2199 &   0008030201/0301/0601 \textbf{0723801101/1201} & 130.2 & 129.7 & 114.1 & 0.0302 & 1.00 & h \\

A\,2597 &   0147330101 \textbf{0723801701} & 110.7 & 112.0 & 85.3 & 0.0852 & 1.11 & h \\

A\,2626 &   0083150201 0148310101 & 50.1 & 50.6 & 41.8 & 0.0573 & 0.84 & h \\

A\,3112 &   0105660101 0603050101/0201 & 186.6 & 190.8 & 153.6 & 0.0750 & 1.13 & h \\

A\,3526 / Centaurus &   0046340101 0406200101 & 151.8 & 153.1 & 128.8 & 0.0103 & 0.83 & h \\

A\,3581 &  0205990101 0504780301/0401  & 113.0 & 117.8 & 84.1 & 0.0214 & 0.72 & c \\

A\,4038 / Klemola 44 & 0204460101  \textbf{0723800801}  & 78.7 & 79.6 & 71.4 & 0.0283 & 0.89 & h \\

A\,4059 &  0109950101/0201 \textbf{0723800901/1001}  & 194.9 & 198.5 & 153.9 & 0.0460 & 0.96 & h \\

AS\,1101 / S\'ersic 159-03 &  0123900101 0147800101  & 121.0 & 122.9 & 108.8 & 0.0580 & 0.98 & h \\

AWM\,7 &  0135950101 0605510101  & 148.7 & 149.6 & 153.9 & 0.0172 & 0.86 & h \\

EXO\,0422 &  0300210401  & 39.5 & 38.9 & 34.9 & 0.0390 & 0.89 & h \\

Fornax / NGC\,1399 &  0012830101 0400620101  & 106.2 & 114.2 & 75.1 & 0.0046 & 0.40 & c \\

HCG\,62 &  0504780501/0601  & 121.8 & 126.7 & 101.6 & 0.0146 & 0.46 & c \\

Hydra\,A &  0109980301 0504260101  & 96.3 & 101.5 & 74.7 & 0.0538 & 1.07 & h \\

M\,49 / NGC\,4472 &  0112550601 0200130101  & 93.1 & 94.8 & 86.3 & 0.0044 & 0.53 & c \\

M\,60 / NGC\,4649 & 0021540201 0502160101  & 118.4 & 119.1 & 108.0 & 0.0037 & 0.53 & c \\

M\,84 / NGC\,4374 & 0673310101  & 32.0 & 34.0 & 30.5 & 0.0034 & 0.46 & c \\

M\,86 / NGC\,4406 &  0108260201  & 68.4 & 70.4 & 47.0 & -0.0009 & 0.49 & c \\

M\,87 / NGC\,4486 &  0114120101 0200920101  & 113.9 & 114.5 & 96.8 & 0.0044 & 0.75 & c \\

M\,89 / NGC\,4552 &  0141570101  & 23.2 & 24.4 & 18.3 & 0.0010 & 0.44 & c \\

MKW\,3s &  0109930101 \textbf{0723801501}  & 147.7 & 148.9 & 126.9 & 0.0450 & 0.95 & h \\

MKW\,4 & 0093060101 \textbf{0723800701}  & 75.5 & 74.9 & 56.9 & 0.0200 & 0.62 & h \\

NGC\,507 & 0080540101 \textbf{0723800301}  & 124.4 & 124.8 & 103.7 & 0.0165 & 0.60 & c \\

NGC\,1316 / Fornax A & 0302780101 0502070201  & 123.7 & 127.2 & 75.2 & 0.0059 & 0.46 & c \\

NGC\,1404 & 0304940101  & 26.8 & 14.8 & 21.0 & 0.0064 & 0.61 & c \\

NGC\,1550 & 0152150101 \textbf{0723800401/0501}  & 166.3 & 167.0 & 128.2 & 0.0123 & 0.62 & c \\

NGC\,3411 & 0146510301  & 21.4 & 21.6 & 19.8 & 0.0155 & 0.47 & c \\

NGC\,4261 & 0056340101 0502120101  & 108.6 & 109.8 & 85.8 & 0.0074 & 0.45 & c \\

NGC\,4325 & 0108860101  & 20.2 & 19.0 & 16.3 & 0.0258 & 0.58 & c \\

NGC\,4636 & 0111190701  & 56.1 & 56.3 & 54.5 & 0.0037 & 0.35 & c \\

NGC\,5044 & 0037950101 0554680101  & 119.0 & 121.8 & 100.4 & 0.0090 & 0.56 & c \\

NGC\,5813 & 0302460101 0554680201/0301  & 138.2 & 143.2 & 106.8 & 0.0064 & 0.44 & c \\

NGC\,5846 & 0021540501 \textbf{0723800101/0201}  & 171.0 & 173.9 & 147.9 & 0.0061 & 0.36 & c \\

Perseus & 0085110101 0305780101  & 155.5 & 156.6 & 132.1 & 0.0183 & 1.29 & h \\

\hline

Total &   & 4\,492.3 & 4\,563.6 & 3\,666.9 &  &  &  \\

\hline                                   
\end{tabular}
\par\end{centering}
$^{(a)}$ Total exposure time after cleaning the data from soft flares (see text).
$^{(b)}$ Redshifts were taken from \citet{2002ApJ...567..716R}, except for A\,189 \citep{2001MNRAS.327..265H}; A\,1991, A\,2626, HCG\,62, and M\,87 \citep[ACCEPT catalog --][]{2009ApJS..182...12C}; M\,89 \citep{2001ApJ...554L.129M}; NGC\,1316 \citep{2014A&A...572L...8P}; NGC\,1404 \citep{2007A&A...476...59M}; M\,84, M\,86, NGC\,4261, and NGC\,4649 \citep{2000MNRAS.313..469S}.
$^{(c)}$ Values of $r_{500}$ were taken from \citet[][and references therein]{2015A&A...575A..38P}.
$^{(d)}$ Classification of the objects. The  letter h stands for the ``hot'' clusters (>1.7 keV), while the  letter c stands for the ``cool'' groups/ellipticals (<1.7 keV). M 87 is an exception, and is classified as cool even though its central temperature is about $\sim$2 keV (see text).\\

\end{table*}




\subsection{Data reduction}

All the data are reduced with the \textit{XMM-Newton} Science Analysis System (SAS) v14.0.0 and by using the calibration files dated by March 2015. The RGS data are the same as used in \citet[][see their Table 1]{2015A&A...575A..38P}, and are reduced the same way. We reduce the EPIC data by using the standard pipeline command \texttt{emproc} and \texttt{epproc}, and, following the standard recommendations, we keep the single to quadruple pixel MOS events (\texttt{PATTERN$\le$12}), and only the single pixel pn events (\texttt{PATTERN$==$0}). Moreover, only the highest quality MOS and pn events (\texttt{FLAG$==$0}) are taken into account. We filter our observations from soft-proton flares by building good time interval (GTI) files following the method described in \citet{2015A&A...575A..37M}. We extract light curves within the 10--12 keV (MOS) and the 12--14 keV (pn) bands in 100 s bins, we calculate the mean count rate $\mu$ and the standard deviation $\sigma$, and we apply a threshold of $\mu \pm 2\sigma$ to the fitted distribution. For safety \citep{2002A&A...389...93L}, we repeat the procedure for the 0.3--10 keV band in 10 s bins. The average fraction of ``good'' time accepted after such a filtering is $\sim$77\%, $\sim$78\%, and $\sim$66\% for MOS\,1, MOS\,2, and pn, respectively, although this fraction  varies widely from pointing to pointing. For each object, the net exposure times of the EPIC instruments are indicated in Table \ref{table:observations}. Combining our whole dataset, we obtain total EPIC and RGS net exposure times of $\sim$4.5 Ms and $\sim$5.1 Ms, respectively.

Finally, point-like sources might pollute our spectra; therefore, we need to discard all of them from the rest of our analysis. We first detect the point sources of every dataset within four spectral bands (0.3--2 keV, 2--4.5 keV, 4.5--7.5 keV, and 7.5--12 keV) using the SAS task \texttt{edetect\_chain}. After a second check by eye, we excise circular regions with 10$''$ of radius around the point sources (except in some specific situations where a larger excision radius is required to remove scattered photons from bright foreground sources). This radius size is estimated to be a good compromise between discarding the polluting flux of the point sources and keeping a maximum of cluster emission in their neighbourhood \citep{2015A&A...575A..37M}. Depending on the target, the typical fraction of removed flux after discarding the point sources varies between $\sim$0.3\% and $\sim$4\%.

\subsection{Spectra extraction}\label{sect:spectra_extraction}

The sources of our sample span a wide range of sizes, masses, and temperatures, and studying their elemental abundances with EPIC over one common astrophysical scale $r_\text{core}$ is difficult in practice. A definition of $r_\text{core}$ as $0.2 r_{500}$\footnote{$r_{500}$ is defined as the radius within which the gas density is 500 times the critical density of the Universe.} for the farther (and, by selection, hotter) clusters is commonly found in the literature \citep[e.g.][]{2007A&A...465..345D}; however,  most of the nearest galaxy groups from our sample are seen at this radius with an angular size $\theta > 15$ arcmin, i.e. beyond the EPIC field of view (FoV). Moreover, extracting spectra over a large $\theta$ for these cooler and more compact objects will include a lot of background, which may dominate beyond $\sim$2--3 keV and entirely flood the K-shell lines of crucial elements such as S, Ar, Ca, or Ni.
Therefore, we choose to split our sample into two subsamples:
\begin{enumerate}

\item The ``hot'' galaxy clusters ($kT \ge 1.7$ keV, 23 objects),

\item The ``cool'' galaxy groups and ellipticals ($kT < 1.7$ keV, 21 objects).

\end{enumerate}
Using the SAS task \texttt{evselect}, we extract the EPIC spectra of every source within a circular region, centred on the peak of the cluster X-ray emission and within a radius of $0.05r_{500}$. Since in the hot clusters a radius of $0.2r_{500}$ can also be reached and provides better signal-to-noise ratio (S/N), we extract these spectra as well.

Table \ref{table:observations} classifies each object in one of these two subsamples. Only M\,87 deviates from  the rule. Indeed, although its main central temperature is about $\sim$2 keV (and is thus considered  a hot object), its $0.2r_{500}$ limit is beyond the EPIC FoV, and only the spectra within $0.05r_{500}$ could be extracted.

We extract the RGS spectra as described in \citet{2015A&A...575A..38P} and in de Plaa et al. (to be submitted). Since the dispersion direction of RGS extends along the whole EPIC FoV, its extraction region will be always different from our circular EPIC extraction regions. Therefore, we extract all the RGS spectra using a cross-dispersion width of 0.8$'$ from the EPIC FoV, which still focuses on the ICM core. In Sect. \ref{sect:systematics} we show that this choice does not affect our results.

The redistribution matrix file (RMF), which gives the channel probability distribution for a photon of given energy, is built using the task \texttt{rmfgen}. The ancillary response file (ARF), which provides the effective area curve as a function of the energy and the position on the detectors, is built using the task \texttt{arfgen}. Both the RMF and the ARF contain all the information relative to the response of the instruments, and need to be further applied for each observation to the spectral modelling\footnote{See the ``Users Guide to the \textit{XMM-Newton} Science Analysis System'', Issue 11.0, 2014 (ESA: \textit{XMM-Newton} SOC).}.


\section{EPIC spectral analysis}\label{sect:spectral_analysis}


We use the SPEX fitting package \citep{1996uxsa.conf..411K} v2.05 to perform the spectral analysis of our sample. We fit all our spectra using the C-statistics \citep[i.e. a modified Cash statistics;][]{1979ApJ...228..939C}, which is appropriate for Poisson-noise dominated data (see the SPEX manual).

Making the reasonable assumption that the hot ICM is in a collisional ionisation equilibrium (CIE) state, throughout this paper we describe its emission using a \texttt{cie} model \citep[based on an updated version of the \texttt{MEKAL} plasma code,][]{1985A&AS...62..197M}. This model includes processes such as collisional ionisation and excitation-autoionisation, as well as radiative and dielectronic recombination (for further details, we refer the reader to the SPEX manual\footnote{\texttt{https://www.sron.nl/spex}}). We adopt the updated ionisation balance calculations of \citet{2009ApJ...691.1540B}. The abundances are calculated from all the transitions and ions of a given element, and are scaled to the proto-solar values\footnote{The proto-solar abundances used in this paper \citep{2009LanB...4B...44L} are the most up-to-date representative abundances of the solar system at its formation, as they are based on meteoritic compositions.} of \citet{2009LanB...4B...44L}.

We fit the cluster emission component in EPIC with a multi-temperature \texttt{cie} model,  the Gaussian Differential Emission Measure (\texttt{gdem}) model, which reproduces a Gaussian temperature distribution in the form
\begin{equation} \label{eq:gdem}
Y(x) = \frac{Y_0}{\sigma_{T} \sqrt{2 \pi}} \exp(\frac{(x-x_\text{mean})^2}{2 \sigma^2_{T}})
,\end{equation}
where $x=\log (kT)$ and $x_\text{mean}=\log (kT_\text{mean})$, $kT_\text{mean}$ is the peak temperature of the distribution, and $\sigma_{T}$ is the full width at half maximum of the distribution \citep[see][]{2006A&A...452..397D}. By definition, $\sigma_T$=0 provides a single-temperature \texttt{cie} model (1T).

The use of a multi-temperature model for such a study is crucial since most of the clusters and groups have a complicated thermal structure in their cores where the cooling rate and  temperature gradient are quite important. Therefore, assuming the plasma to be isothermal in general may lead to the so-called Fe bias, i.e. an underestimate of the Fe abundance \citep[see e.g.][]{1994ApJ...427...86B,1998MNRAS.296..977B,2000MNRAS.311..176B}. The effects of different thermal models on the abundances and a comparison between EPIC and RGS measurements are  discussed below (Sect. \ref{sect:systematics}).

For both EPIC and RGS abundances, we also correct the O and Ne estimates from updated calculations of the radiative recombination contribution to the cluster emission as a function of its mean temperature. We do so by multiplying the O and Ne best-fit measurements of each object by the factors of the corrected \ion{O}{viii} and \ion{Ne}{x} Lyman $\alpha$ fluxes, as described in Appendix \ref{sect:RR_rates}. On average, these corrections increase the O and Ne abundances by $\sim$20\% and $\sim$6\%, respectively.

The Galactic absorption that we apply to our fitted thermal models is modelled by the transmission of a neutral plasma (\texttt{hot} model, for which $kT$=0.5 eV). In EPIC, the hydrogen column density $N_\text{H}$ has been estimated from a grid search of (fixed) values within 
\begin{equation}
N_\ion{H}{i} - 5\times 10^{19} \text{ cm}^{-2} \le N_\text{H} \le N_\text{H,tot} + 1 \times 10^{20} \text{ cm}^{-2},
\end{equation}
where $N_\ion{H}{i}$ and $N_\text{H,tot}$ are respectively the neutral \citep{2005A&A...440..775K} and total (neutral and molecular) hydrogen column densities estimated using the method of \citet{2013MNRAS.431..394W}. More details on the reasons for this approach are given in Appendix \ref{sect:N_H}.

Details on the RGS spectral analysis (including models, free parameters, and background treatment) can be found in de Plaa et al. (to be submitted), and in \citet{2015A&A...575A..38P}.

\subsection{Background modelling}\label{subsect:bkg_modelling}

Although the clusters considered in this work are usually bright and display a high S/N within their core, in most of them the EPIC background can still play a significant role, especially in the hard spectral bands (i.e. $\ga$2 keV) where less thermal emission is expected. Because a slightly incorrect scaling in the subtraction of background data (taken from either filter closed wheel data or blank sky observations) can  significantly affect the temperature estimates and thus bias the spectral analyses \citep{2006A&A...452..397D}, we choose here to model the background directly in our spectra. The method we use is extensively described in \citet{2015A&A...575A..37M}. In summary, we model five separate background components:
\begin{itemize}
\item the local hot bubble, modelled by a non-absorbed isothermal \texttt{cie} component, whose abundances are kept proto-solar;
\item the galactic thermal emission, modelled by an absorbed isothermal \texttt{cie} component, whose abundances are also kept proto-solar;
\item the unresolved point sources, modelled by an absorbed power law with a photon index fixed to 1.41 \citep{2004A&A...419..837D};
\item the hard particle background, modelled by a broken power law (unfolded by the effective area) and several instrumental Gaussian profiles. The parameters are taken from \citet{2015A&A...575A..37M}, except for all the normalisations, which are left free;
\item the soft-proton background, modelled by a power law (unfolded by the effective area). The parameters are estimated using EPIC spectra covering the total FoV (where the parameters of the \texttt{gdem} component from the ICM are also left free; see Sect. \ref{sect:global_fits}). From such spectra, the ICM emission can be easily constrained in the soft band ($\la$2 keV), while the particle backgrounds clearly dominate the harder bands, making the soft-proton background contribution easier to estimate.
\end{itemize}

The fluxes and the temperatures of the local hot bubble and galactic thermal emission components, as well as the flux of the unresolved point sources, are estimated from \textit{ROSAT} PSPC spectra extracted from the region beyond $r_{200}$ of each object \citep{2009ApJ...699.1178Z}.

Finally, all these background components are fixed and rescaled to the sky area of our EPIC core spectra (except the normalisation of the hard particle background, which we always left free in order to avoid incorrect scalings and temperature biases, see above).

In addition to the background described above, M\,87, M\,89, NGC\,4261, NGC\,4636, and NGC\,5813 host a powerful active galactic nucleus (AGN), which can generate cavities in the hot gas \citep[e.g.][]{2013MNRAS.432..530R}, but can also pollute the total X-ray emission. For each of these observations, we start by extracting a circular region of 30$''$ centred on the AGN, and we fit its EPIC spectra with an absorbed power law (in addition to the cluster emission and the background components described above). We then extrapolate this additional component to the EPIC core spectra, fixing its column density and photon index values derived from the 30$''$ aperture region, and rescaling its normalisation to the area ratio of these two extracted regions. We note that in the EPIC core spectra the 0.5--10 keV flux of the AGN component is never larger than $\sim$15--20\% of the cluster emission. This justifies {a posteriori} our choice of fitting the AGN contribution rather than excising the AGN,  i.e. where the peak of the cluster emission is.

\subsection{Global fits}\label{sect:global_fits}

In addition to the normalisation of the hard particle background, the only parameters that are left free when fitting our spectral components to the EPIC spectra of the core regions are the normalisation (or emission measure, $Y = \int n_e n_\text{H} dV$) of the \texttt{gdem} model; its mean temperature ($kT_\text{mean}$);  $\sigma_T$; and the abundances of O, Ne, Mg, Si, S, Ar, Ca, Fe, and Ni. The other $Z \ge 6$ abundance parameters are coupled to the value of the Fe abundance parameter, and are thus not free. The MOS and pn spectra of every pointing are fitted simultaneously. Since the large number of total parameters prevents us from fitting simultaneously three observations or more, for each object we form pairs of two pointings that we fit simultaneously. For objects including $\ge$3 pointings, we then combine the results of the fitted pairs using a weighting factor of $1/\sigma_i^2$, where $\sigma_i$ is the statistical error on the considered parameter.

An important variable that might affect our EPIC results is the spectral ranges of our fits. In particular, significant cross-calibration issues between MOS and pn have been reported in the soft bands \citep{2014A&A...564A..75R,2015A&A...575A..30S}. Similarly, we observe a sharp and extremely variable soft tail in the EPIC filter wheel closed events\footnote{See also the XMM-Newton Calibration Technical Note, XMM-SOC-CAL-TN-0018 (Ed: Guainazzi, 2014).} that might considerably affect the spectra below 0.5 keV. On the other hand, we would like to keep our spectral range as large as possible, for instance to estimate the abundance measurements of O and the temperature structure in the Fe-L complex. A good compromise is found by using the 0.5--10 keV and 0.6--10 keV bands for MOS and pn, respectively.



\subsection{Local fits}\label{sect:local_fits}

In the case of a plasma in CIE, the abundances of a given element showing prominent and well-resolved emission lines are easy to derive as they are proportional to the ratio between the line flux and the continuum flux, namely the equivalent width (EW). However, as a consequence of the imperfections of the EPIC instruments effective areas, when fitting the EPIC spectra over a large range (0.5/0.6--10 keV in the previous subsection, hereafter the ``global'' fits), the modelled continuum emission may be slightly over- or underestimated in some specific energy bands. Consequently, the modelled line fluxes tend to compensate the continuum discrepancies in the global  fits, by, in turn, slightly under- or overestimating the value of the abundance parameters.

This effect, already discussed in \cite{2015A&A...575A..37M}, can be easily corrected by fitting the EPIC spectra locally, in order to allow the modelled continuum to be fitted to its correct (local) level. Therefore, we re-fit the EPIC spectra within local bands successively centred around the strongest K-shell lines of each element (except Ne, whose strongest lines reside in the Fe-L complex and are not resolved by the EPIC instruments). The temperature parameters $kT_\text{mean}$ and $\sigma_T$ are frozen to their EPIC global best-fit values, in such a way that in every local fit, the free parameters are only the (local) normalisation $Y$ and the abundance of the considered element. We compare these abundances estimated locally in MOS (MOS\,1 and MOS\,2 are fitted simultaneously) and in pn individually. If the MOS and pn abundances agree within 1$\sigma$, we combine the measurements using a weighting factor of $1/\sigma_i^2$ (see also Sect. \ref{sect:global_fits}). Otherwise, we compute the weighted average and artificially increase the combined uncertainties until they fully cover the extreme MOS and pn 1$\sigma$ values. By applying such a conservative method to each object, we cover individual systematic uncertainties related to the EPIC cross-calibration issues (see also Sect. \ref{sect:systematics}), and ensure getting fully reliable abundance measurements.

Hereafter, all the EPIC abundances are locally corrected, unless otherwise stated. We note, however, that the EPIC Fe abundances reported in this paper are obtained using global fits because they are more accurately determined using both the Fe-K and Fe-L complexes. Except A\,3526 (for which we estimate Fe using local fits), all the other objects show (<2$\sigma$) consistent EPIC Fe abundances when using successively local and global fits, so this choice does not affect our results.


\section{Results}\label{sect:results}

\begin{figure*}[!]
        \centering
                \includegraphics[width=0.75\textwidth]{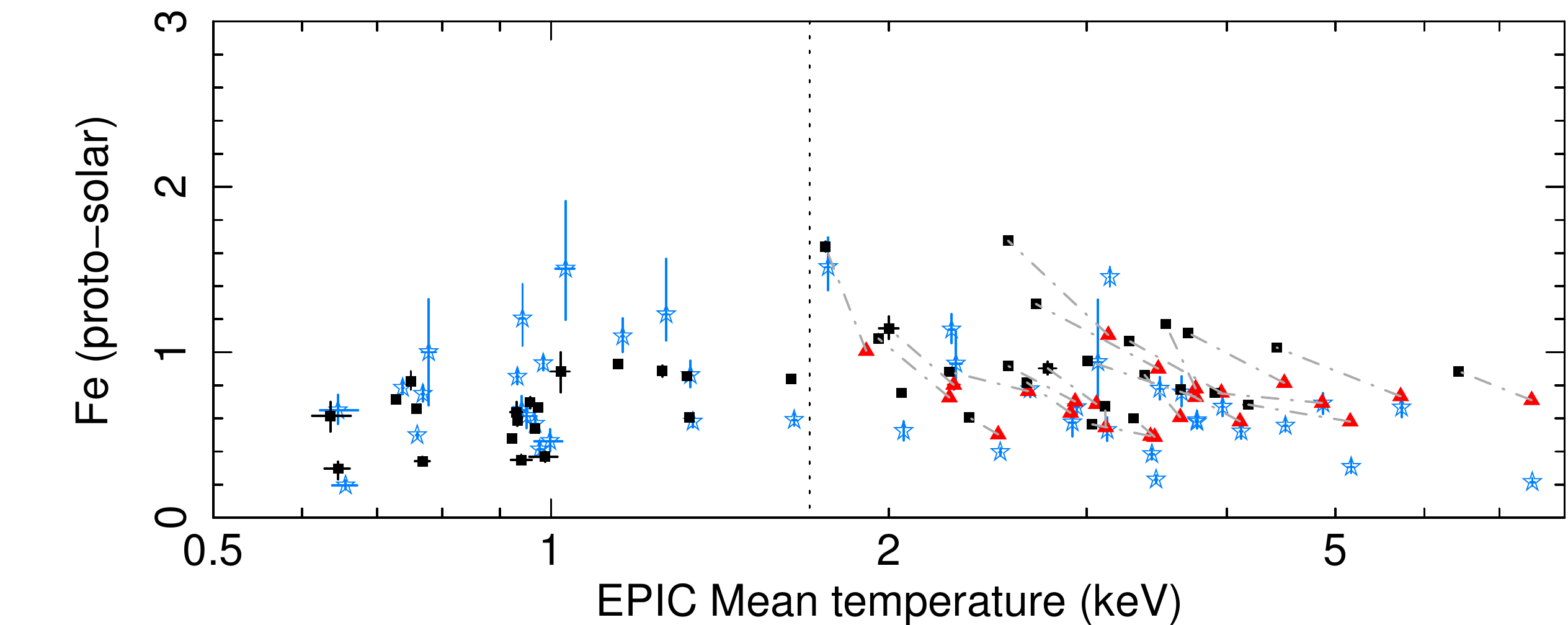}

        \caption{Mean temperature (EPIC) versus absolute Fe abundance for the full sample. The black squares and the red triangles show the EPIC measurements within $0.05r_{500}$ and $0.2r_{500}$, respectively. Each pair of measurements ($0.05r_{500}$,$0.2r_{500}$) that belong to the same hot cluster is connected by a grey dash-dotted line. The blue stars show the RGS measurements (adapted from de Plaa et al., to be submitted), scaled on their respective EPIC mean temperature within $0.2r_{500}$. The vertical black dotted line separates the cool groups/ellipticals from the hot clusters (see text).}
\label{fig:sample_kT_abun_Fe}
\end{figure*}
\begin{figure*}[!]
        \centering
                \includegraphics[width=0.49\textwidth]{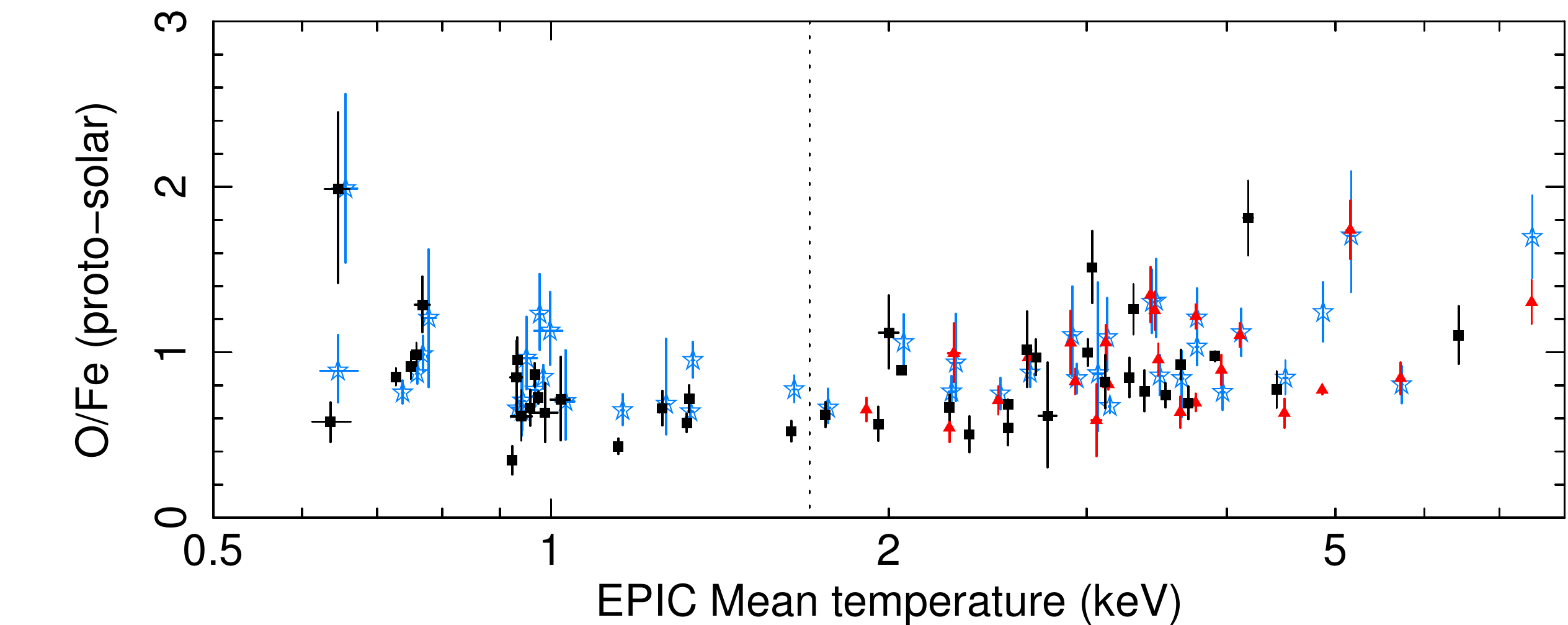}
                \includegraphics[width=0.49\textwidth]{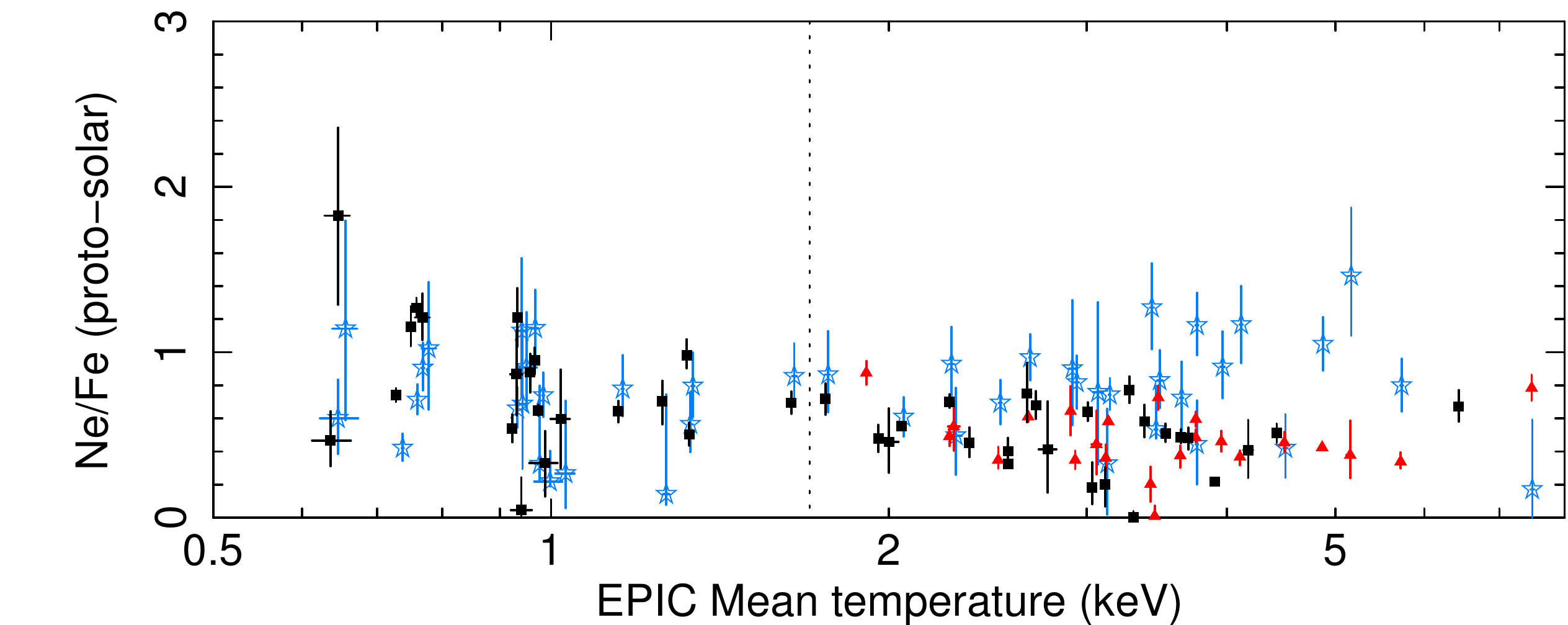}
\\
                \includegraphics[width=0.49\textwidth]{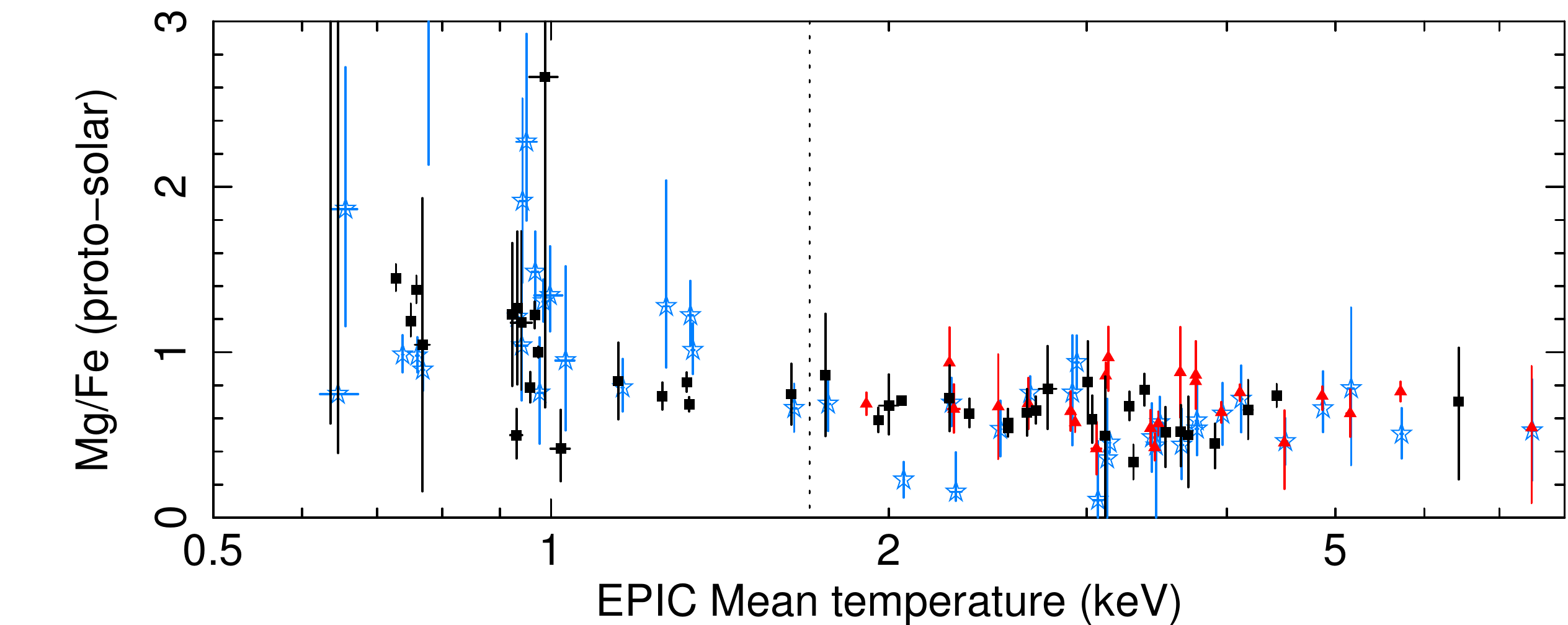}
                \includegraphics[width=0.49\textwidth]{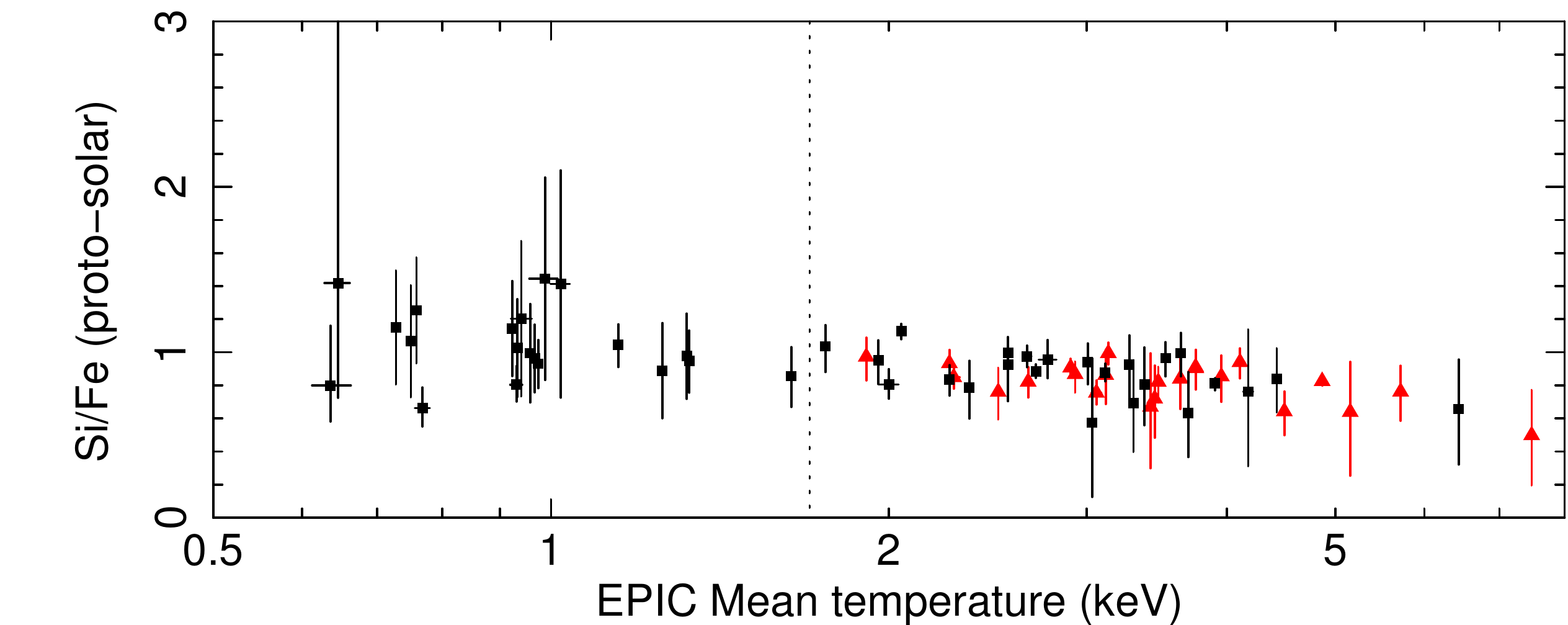}
\\
                \includegraphics[width=0.49\textwidth]{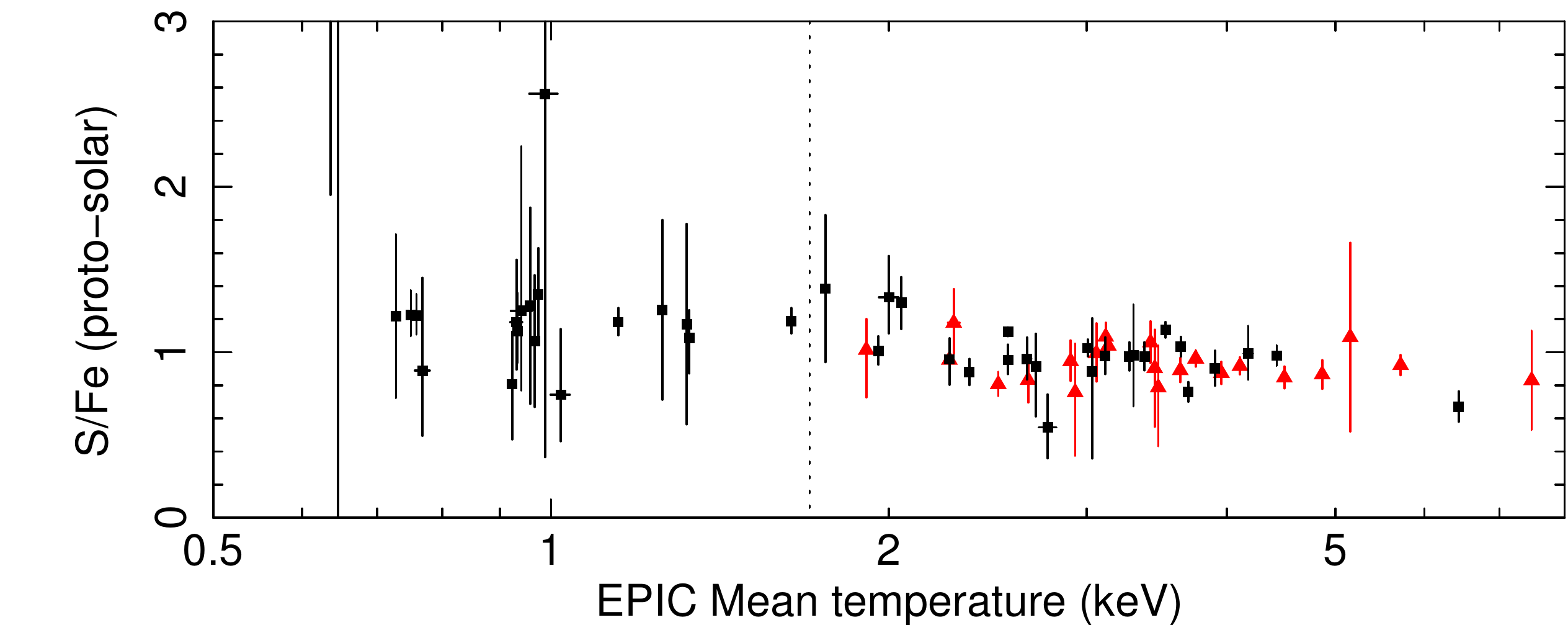}
                \includegraphics[width=0.49\textwidth]{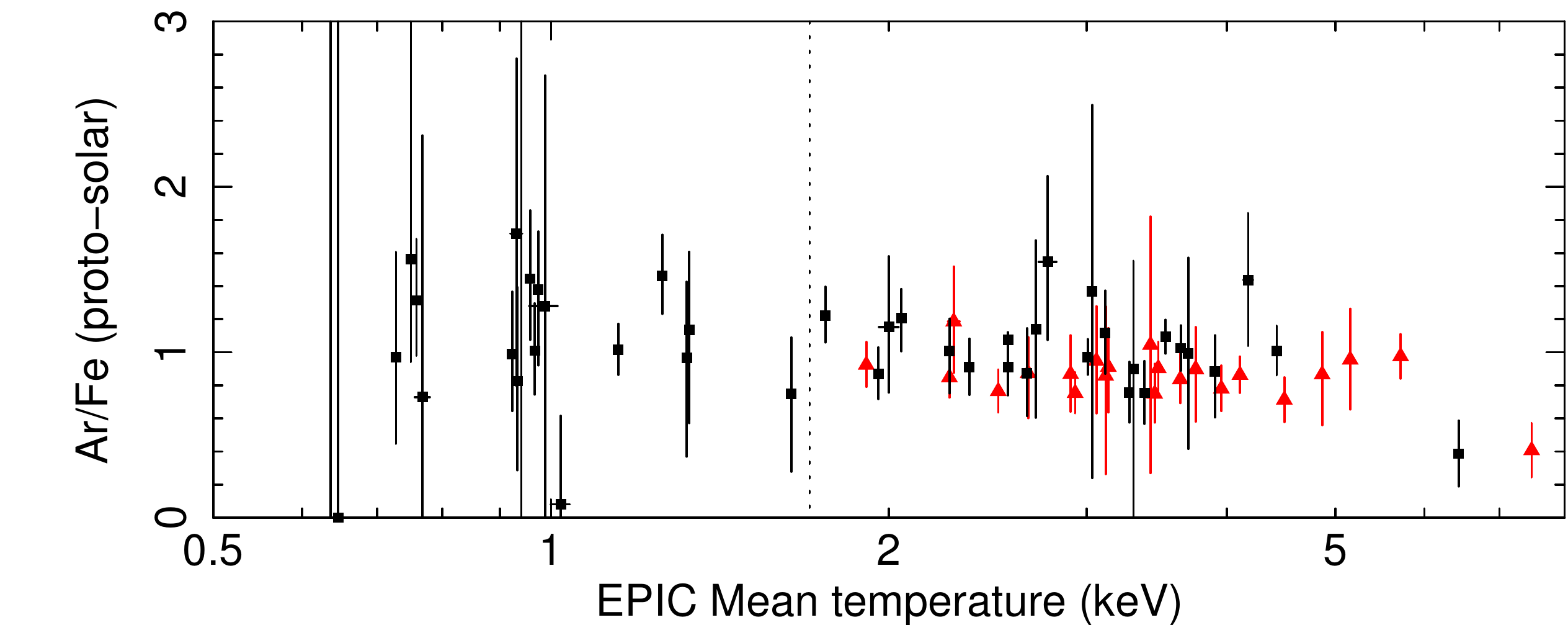}
\\
                \includegraphics[width=0.49\textwidth]{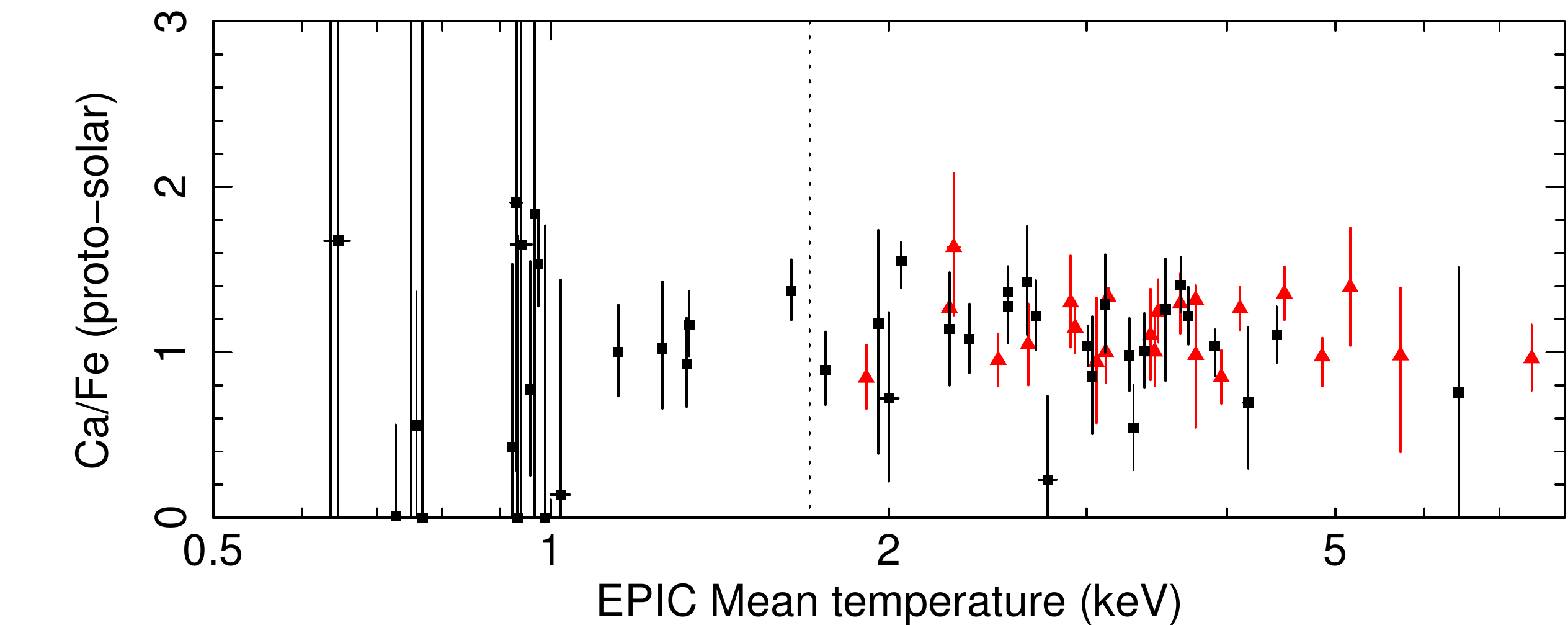}
                \includegraphics[width=0.49\textwidth]{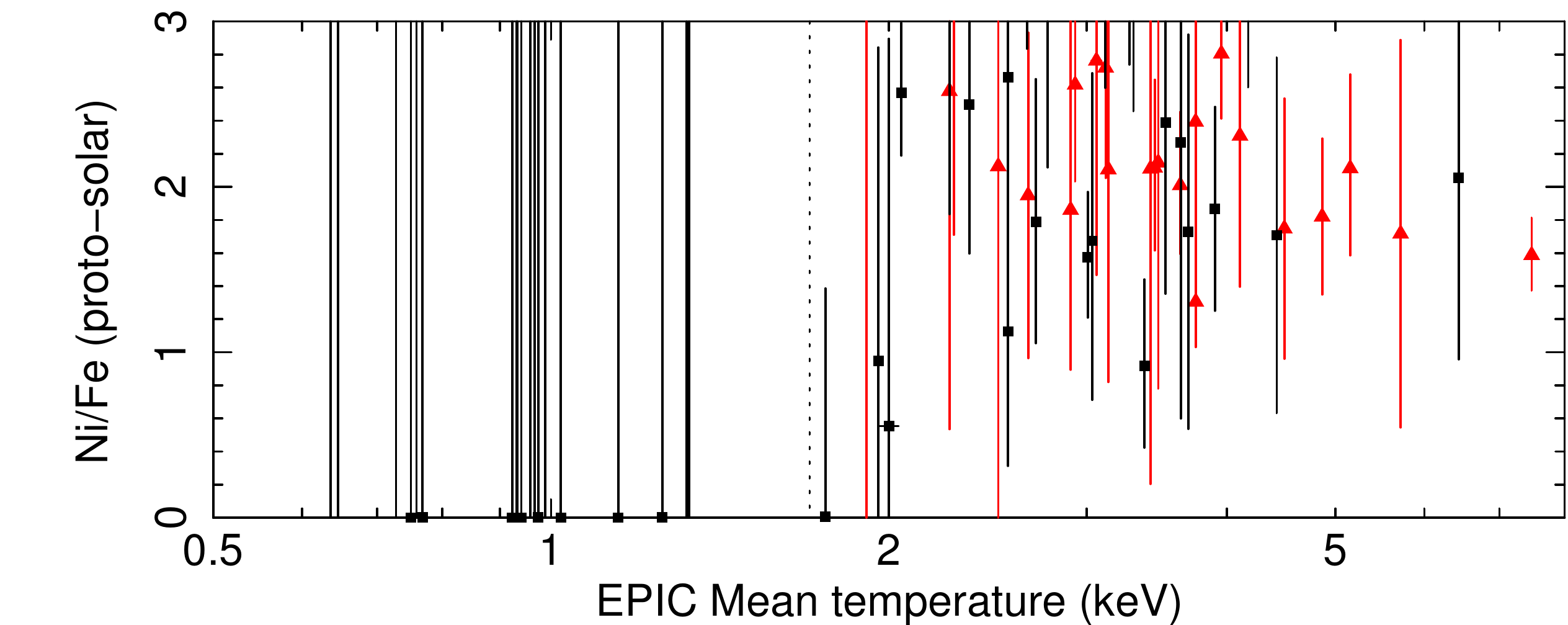}

        \caption{Same as Fig. \ref{fig:sample_kT_abun_Fe} for the other (relative-to-Fe) abundance ratios. For clarity, the ($0.05r_{500}$,$0.2r_{500}$) pairs are not shown explicitly.}
\label{fig:sample_kT_abun_others}
\end{figure*}

The final abundance
estimates for EPIC (within $0.2r_{500}$ and $0.05r_{500}$) and RGS  of all the objects in the sample are presented in Fig. \ref{fig:sample_kT_abun_Fe} (Fe abundance) and Fig. \ref{fig:sample_kT_abun_others} (other relative-to-Fe abundance ratios), spread over their EPIC mean temperatures.

As can be seen in Fig. \ref{fig:sample_kT_abun_Fe}, some sources show a significant discrepancy between their EPIC and RGS measured Fe abundances. This is not surprising, since the RGS extraction regions always have the same angular size ($\sim$30$'$$\times$0.8$'$), while the radius of the circular EPIC extraction regions is different for each object (Sect. \ref{sect:spectra_extraction}). Moreover, owing to its poorly constrained continuum level and its limited spectral range (in particular with no access to the Fe-K lines), RGS is not very suitable for deriving absolute Fe abundances. In the case of very extended sources, the instrumental line broadening makes Fe even more difficult to derive with RGS, leading to larger uncertainties. Nevertheless, the relative abundance ratios O/Fe, Ne/Fe, and Mg/Fe measured with RGS do not depend on the continuum and are easier to constrain (Sect. \ref{sect:select_abundances}).
 
\begin{figure}[!]
        \centering
                \includegraphics[width=0.49\textwidth]{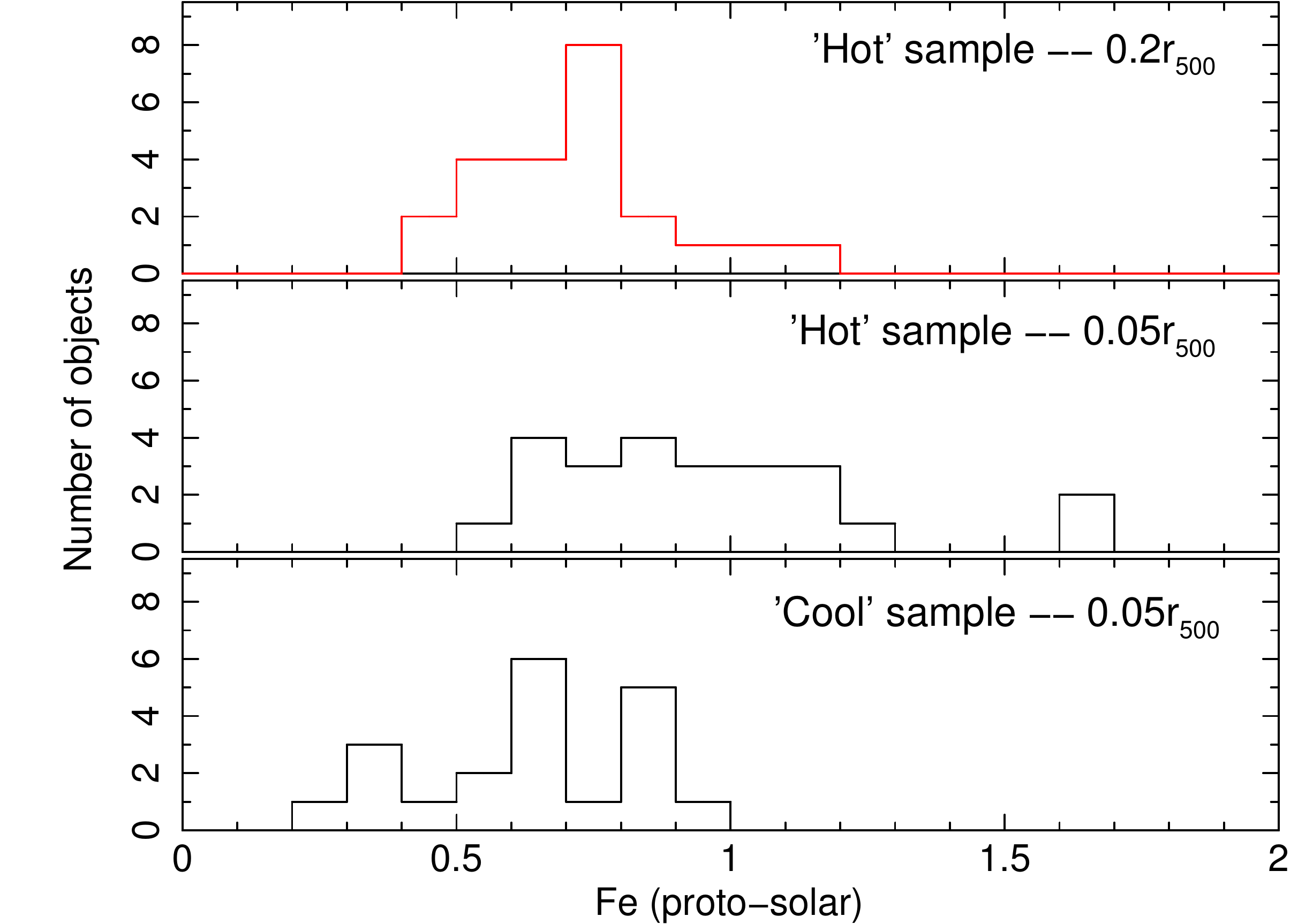}

        \caption{Distributions of the EPIC absolute Fe abundances for all the objects in our sample. Three subsamples (hot clusters within $0.2r_{500}$ and $0.05r_{500}$, and cool groups within $0.05r_{500}$) are considered separately (see also Fig. \ref{fig:sample_kT_abun_Fe}).}
\label{fig:sample_histo_Fe_best}
\end{figure}

Within $0.2r_{500}$ (red triangles), the Fe abundance of the hot clusters are somewhat dispersed, with a mean value of 0.71. Within $0.05r_{500}$ (black squares), the mean Fe abundance in the cool groups/ellipticals is 0.64, while in the hot clusters it is 0.78 (see also Fig. \ref{fig:sample_histo_Fe_best}). We estimate the intrinsic scatter in our subsamples, and its upper and lower 1$\sigma$ limits by following the method described in \citet{2007A&A...465..345D}. Knowing the statistical errors $\sigma_\text{stat}$ of our measurements, we determine the intrinsic scatter $\sigma_\text{int}$ using fits to our data with a constant model and with total uncertainties $\sqrt{ \sigma_\text{stat}^{2} + \sigma_\text{int}^{2} }$ that have $\chi^2 = k \pm \sqrt{2k}$ (where $k$ is the number of degrees of freedom). For the hot clusters we find an intrinsic scatter of $(21 \pm 4)\%$ within $0.2r_{500}$, and $(33 \pm 7) \%$ within $0.05r_{500}$. The intrinsic scatter in the cool groups ($0.05r_{500}$) is  $(31 \pm 5)\%$, which is comparable to the value found in hot clusters within the same core radius.
Finally, we note an interesting trend regarding the pair of measurements ($0.05r_{500}$,$0.2r_{500}$) for each hot cluster (blue dotted lines). When the cluster mean temperature increases, the temperature gradient seems to increase, while on the contrary, the Fe gradient seems to flatten.

All the abundance ratios shown in Fig. \ref{fig:sample_kT_abun_others} are consistent with being uniform over the considered temperatures range, even when considering the two different EPIC extraction regions. This is particularly striking for Si/Fe (although a slightly decreasing trend cannot be excluded) and S/Fe. This trend is investigated more quantitatively in Sect. \ref{sect:systematics} where we compare the average abundance ratios of the hot and the cool objects. Moreover, we note that both EPIC and RGS measurements are consistent; the  exception is  Ne/Fe, for which the RGS measurements remain uniform while the EPIC values suggest a decrease with temperature (see discussion in Sect. \ref{sect:select_abundances}, and a further inspection in Sect. \ref{sect:systematics}). Finally, although their uncertainties are large and deriving any trend is very difficult, we note that the Ni/Fe abundance ratios are all consistent with being larger than the proto-solar value.

\subsection{Estimating reliable average abundances}\label{sect:select_abundances}

Assuming that all these relative-to-Fe abundance ratios are indeed uniform over clusters and do not  depend (much) on their histories, we can combine our individual measurements and estimate for each element one average abundance ratio representative of the nearby cool-core ICM as a whole. We estimate the average relative-to-Fe abundance of a given element ``X'' by using the weighting factors $1/\sigma(\text{X/Fe})_i^2$, where $\sigma(\text{X/Fe})_i$ is the uncertainty on the X/Fe abundance in the $i$th observation. In the case of asymmetric X/Fe uncertainties in some observations, we systematically choose the larger one (in absolute value).

In addition to studying the subsamples of the hot clusters (within either $0.2r_{500}$ or $0.05r_{500}$), we can also combine the hot subsample (within $0.2r_{500}$) with the cool subsample (within $0.05r_{500}$), in order to get a ``full'' sample, named hereafter $(0.05+0.2)r_{500}$, which contains the highest statistics. A complete comparison of this full sample with the three  subsamples mentioned above is discussed in Sect. \ref{sect:systematics}.

As mentioned earlier, RGS measures the absolute Fe abundance with a high degree of uncertainty. However, it is quite reliable in measuring the abundance ratios of O/Fe, Ne/Fe, and sometimes Mg/Fe (assuming a low redshift and a high S/N, which is the case for our sample).
Unlike RGS, EPIC is not very suitable for measuring O/Fe abundance ratios (whose main emission lines reside at $\sim$0.6 keV near the O absorption edge and where the calibration is somewhat uncertain) and Ne/Fe abundance ratios (whose K-shell transitions are within the Fe-L complex, which depends on the temperature structure and is not resolved by the EPIC instruments), but can in principle make reliable measurements of all the other considered ones. Moreover, EPIC observes both the Fe-L and Fe-K complexes, as well as the continuum emission, and thus provides more trustworthy absolute Fe abundances and temperatures. We note that we find large positive residuals around 1.2 keV in the EPIC spectra of NGC\,5813 and NGC\,5846, which prevents us from estimating reasonable Mg abundances, even by performing local fits. For these two groups, the Mg/Fe ratios inferred from RGS are undoubtedly more reliable.

Taking these instrumental characteristics into account, in the following we use the O/Fe and Ne/Fe abundances from RGS. We use EPIC for the Mg/Fe (except in NGC\,5813 and NGC\,5846), Si/Fe, S/Fe, Ar/Fe, Ca/Fe, Fe, and Ni/Fe abundances. We discuss more extensively the robustness of this choice in Sect. \ref{sect:systematics}. Table \ref{table:full_results} shows the best estimated temperature and selected abundance measurements for all the objects in our samples. The average abundance ratios and their statistical uncertainties $\sigma_\text{stat}$ are indicated in the second and third columns of Table \ref{table:systematics}. We note again that O/Fe and Ne/Fe have been corrected from updated radiative recombination calculations (Appendix \ref{sect:RR_rates}).

\subsection{EPIC stacked residuals}\label{sect:residuals}

\begin{figure}[!]
        \centering
                \includegraphics[width=0.495\textwidth]{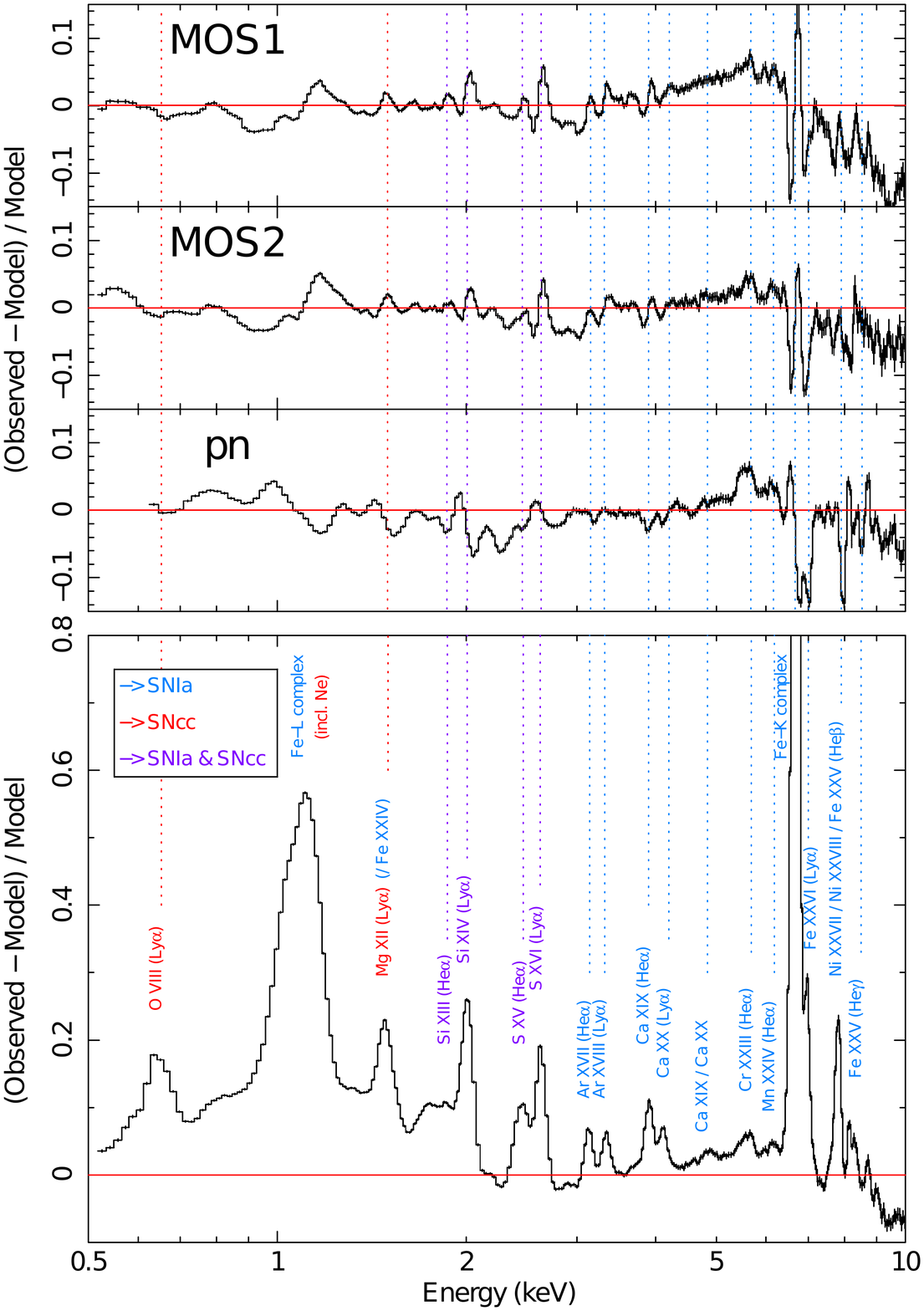}

        \caption{\textit{Top}: EPIC MOS\,1, MOS\,2, and pn stacked and redshift-corrected residuals of the $(0.05+0.2)r_{500}$ sample (using a \texttt{gdem} model). Before stacking, the MOS and pn spectra of every pointing were fitted simultaneously with coupled parameters. The vertical dotted lines indicate the position of the detected line emissions in the EPIC spectra (see lower panel). \textit{Bottom}: EPIC stacked and redshift-corrected residuals of the $(0.05+0.2)r_{500}$ sample (using a \texttt{gdem} model, all instruments combined). Before stacking, the MOS and pn spectra of every pointing were fitted independently and the line emission was set to zero in the model. The height of the peak of the Fe-K complex is $\sim$1.95.}
\label{fig:residuals}
\end{figure}

The large net exposure time allows us to stack the residuals of the previously fitted global EPIC spectra. The residuals of each observation are obtained after  fitting the three instruments simultaneously for each pointing (Sect. \ref{sect:global_fits}), and are corrected from their respective redshift before the stacking process. The residuals are summed over observations following
\begin{equation}
\sum_{i=1}^{N} \sum_{k=1}^{M} w_{i,k} \, (\text{data}(k)_i/\text{model}(k)_i - 1),
\end{equation} 
where $\text{data}(k)_i$ and $\text{model}(k)_i$ are respectively the measured and modelled count rates of the $i$th observation at its $k$th spectral bin, $N$ is the total number of observations, $M$ is the number of spectral bins (in the considered spectrum), and $w_{i,k}$ is the weight used to stack the results. This weight, which depends on both the observation and the spectral bin considered, is the product of two values: the inverse square of the statistical error of $\text{data}(k)_i$ and a factor, between 0 and 1, corresponding to the overlapping fraction between a bin from a reference spectrum, and a bin from a spectrum to be stacked to the reference one \citep[e.g. if the ``reference spectrum'' and ``stacking spectrum'' bins do not overlap, the overlapping fraction is 0, if they fully overlap, the fraction is 1; see also][]{2008A&A...487..461L}. This second factor is necessary because the spectra (or spectral residuals) from different observations have different offsets in their rest frame binning  owing to their different redshift corrections. Figure \ref{fig:residuals} (top three panels) shows the stacked MOS\,1, MOS\,2, and pn residuals.

Although the deviations are not larger than a few per cent, remaining cross-calibration issues between MOS and pn effective areas clearly appear, and positive residuals in one instrument are often compensated by negative residuals in the other, especially around the Fe-L complex (and more generally below 2 keV). In the 4--6 keV band, the model underestimates the spectra, while above 7 keV, the opposite situation occurs. It is also worth mentioning the apparent slightly overestimated broadening of the modelled Fe-K line complex, in particular in MOS (seen through the characteristic dips on both sides of the peak), which is likely due to imperfections of the RMF. The pn stacked residuals also suggest a small offset due to incorrect energy calibration. Although the last two points should not significantly affect our results, the overall shape of the stacked residuals clearly illustrates that, despite past and recent efforts to cross-calibrate the EPIC instruments, imperfections are still present and bring additional uncertainties in the parameter determination \citep[see also][]{2015A&A...575A..30S}. In particular, the biased determination of the continuum, especially beyond 4 keV, emphasises the importance of using a local fitting method to derive reliable abundances (Sect. \ref{sect:local_fits}).

We do the same exercise, this time by fitting the instruments independently (to minimise the cross-calibration residuals discussed above), and by setting all the line emission to zero in the \texttt{gdem} model after having fitted the spectra. We calculate the residuals relative to this ``continuum only'' model for each observation, and we sum the residuals as described above. The stacked result is shown in the lower panel of Fig. \ref{fig:residuals}, and reveals all the emission lines/complexes that the EPIC instruments are able to resolve. The small stacked error bars on the residuals allow us to detect the main emission lines of chromium (Cr) and manganese (Mn) around $\sim$5.7 keV and $\sim$6.2 keV respectively. Following the Gauss method described in \citet{2015A&A...575A..37M}, we re-fit locally the EPIC spectra of every pointing with a local continuum and an additional Gaussian centred successively on these two energies.  From this we get the EWs of the two lines, which we can convert into Cr and Mn abundances. After stacking these measurements over the whole sample, the MOS and pn instruments find a positive detection of Cr/Fe with >7$\sigma$ and >4$\sigma$ significances, respectively. For Mn/Fe, the positive detection is >5$\sigma$ in both MOS and pn. Combining the MOS and pn instruments, we obtain average Cr/Fe and Mn/Fe abundances of $1.56 \pm 0.19$ and $1.70 \pm 0.22$, respectively. These abundances are not so different from the Fe values assumed for Cr and Mn in the previously discussed fits (Sect. \ref{sect:global_fits}), and consequently, their residuals did not bias our fits much, if at all. We also note that because the error bars of these abundances in individual pointings are often 1$\sigma$ consistent with zero, we must ensure that negative abundances are allowed in order to avoid statistical biases when averaging over the whole sample \citep{2008A&A...487..461L}.

\begin{figure}[!]
        \centering
                \includegraphics[width=0.495\textwidth]{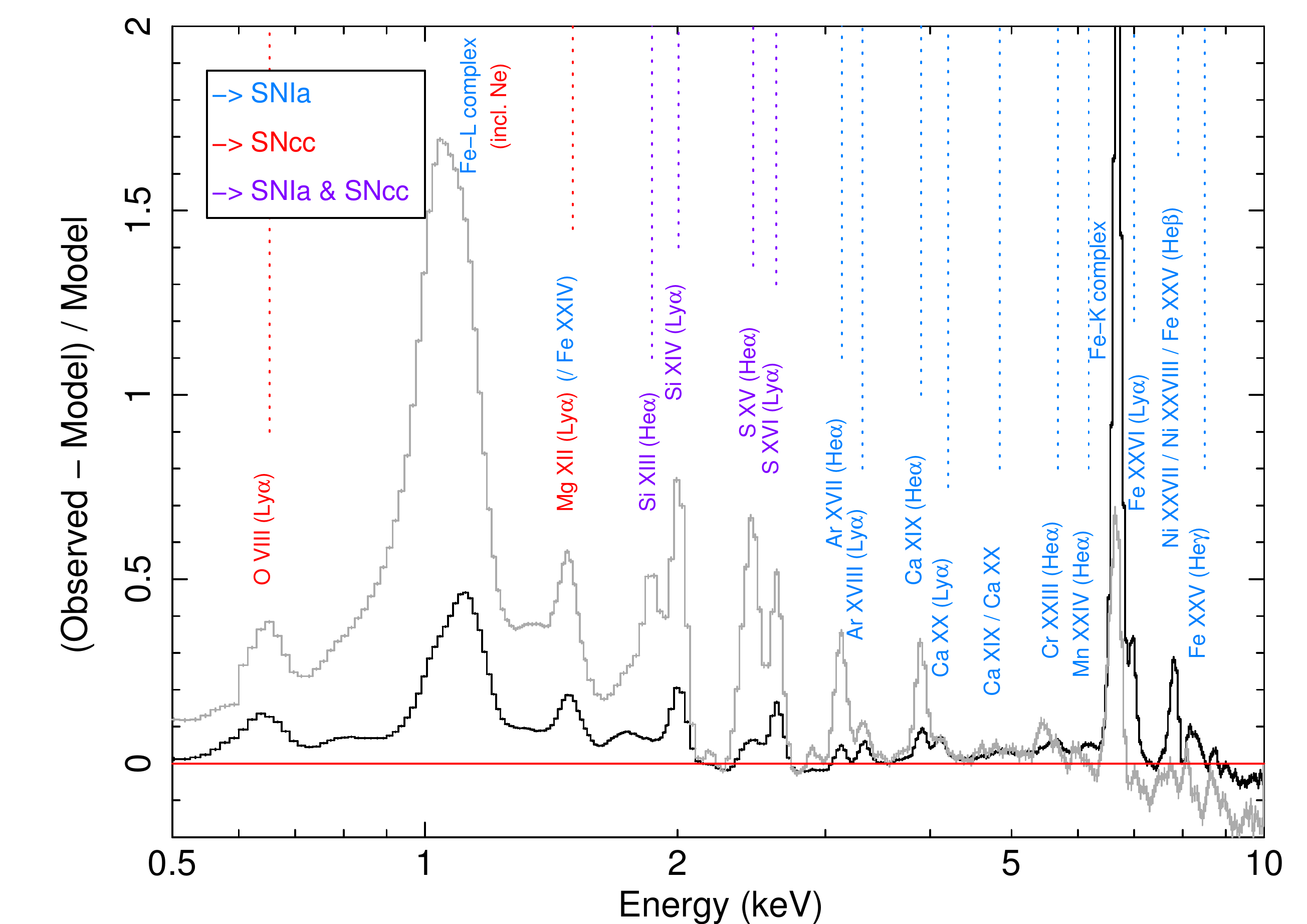}

        \caption{Same as Fig. \ref{fig:residuals} (bottom), this time comparing the $0.05r_{500}$ cool (grey curve) and the $0.2r_{500}$ hot (black curve) subsamples. The height of the peak of the Fe-K complex in the $0.2r_{500}$ hot sample is $\sim$2.23.}
\label{fig:residuals_hotcold}
\end{figure}

The stacking process described above can also be performed separately in the hot and cool subsamples, respectively within $0.2r_{500}$ and $0.05r_{500}$. This comparison is shown in Fig. \ref{fig:residuals_hotcold}. Unsurprisingly, most of the emission lines, including the Fe-L complex, are clearly enhanced in the cool subsample (grey curve), while the Fe-K and Ni-K lines are more prominent in the hot subsample (black). Beyond $\sim$6 keV, the overestimate of the continuum (discussed above and in Sect \ref{sect:local_fits}) also seems  more important in the cool subsample.

\subsection{Systematic uncertainties}\label{sect:systematics}

A crucial point when averaging the abundances over a large sample is that the stacked statistical uncertainties become very small. Therefore, the systematic uncertainties may clearly dominate, and care must be taken to evaluate them properly. The average abundance ratios and all their uncertainties discussed below are summarised in Table \ref{table:systematics}.

First, because of their different chemical histories, it seems reasonable to assume that some clusters intrinsically deviate from the average estimated abundances. Such an intrinsic scatter $\sigma_\text{int}$ has been already introduced, and has been estimated (as well as their 1$\sigma$ uncertainties) for all the available abundance ratios (Table \ref{table:systematics}). Except for O/Fe ($\sim$16\%) and Mg/Fe ($\sim$29\%), the intrinsic scatter of the other elements are of the order of a few percent. In order to remain as conservative as possible in determining our final abundance ratios, we choose to consider the most extreme case where the true instrinsic scatters would actually correspond to the (1$\sigma$) upper limits of $\sigma_\text{int}$. In the following, the systematic uncertainties we associate with the instrinsic scatters are, therefore, $\sigma_\text{int} + 1$$\sigma$. Owing to their still large statistical error bars, no intrinsic scatter was needed for Ar/Fe, Cr/Fe, Mn/Fe, and Ni/Fe.

\begin{figure*}[!]
        \centering
                \includegraphics[width=0.49\textwidth]{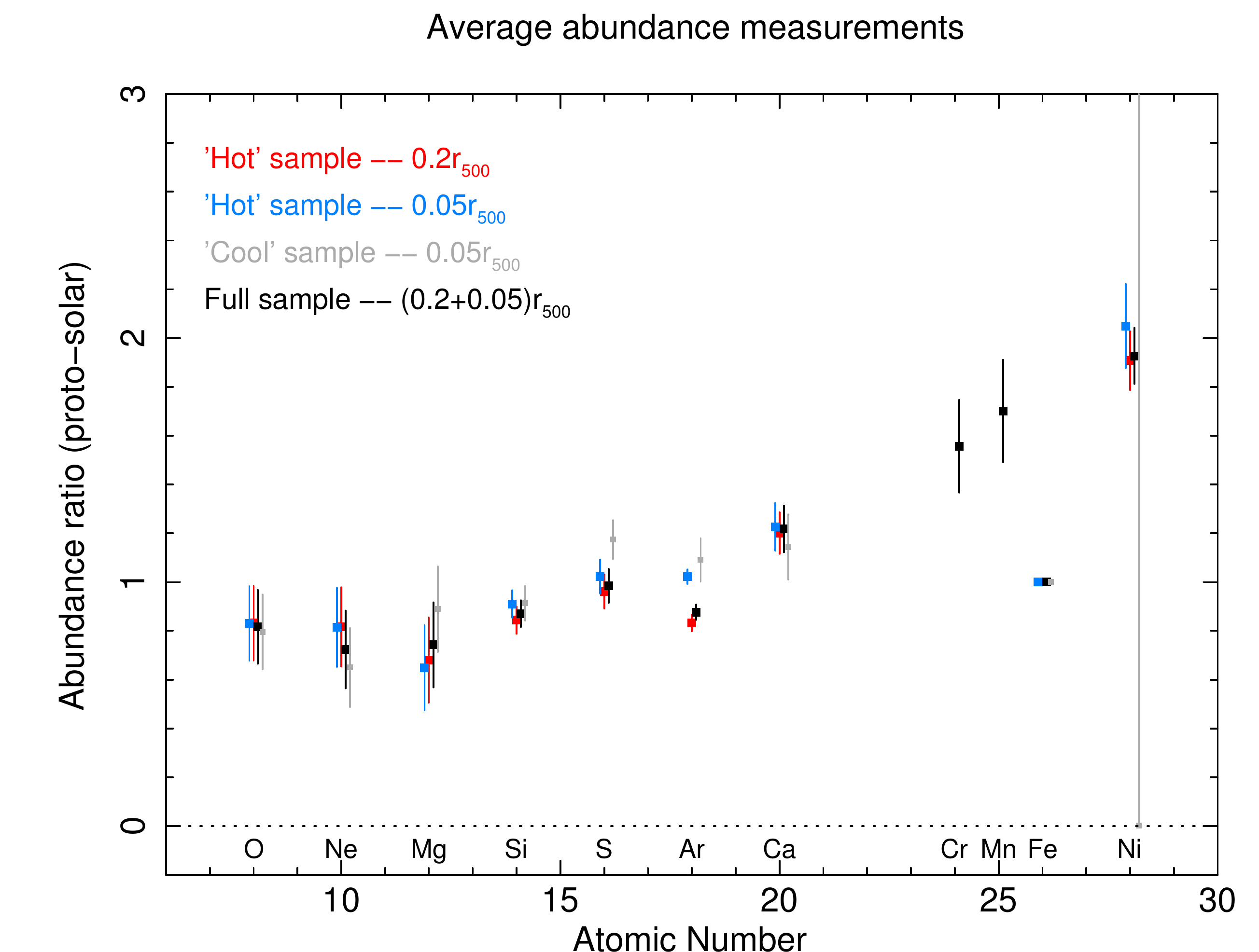}
                \includegraphics[width=0.49\textwidth]{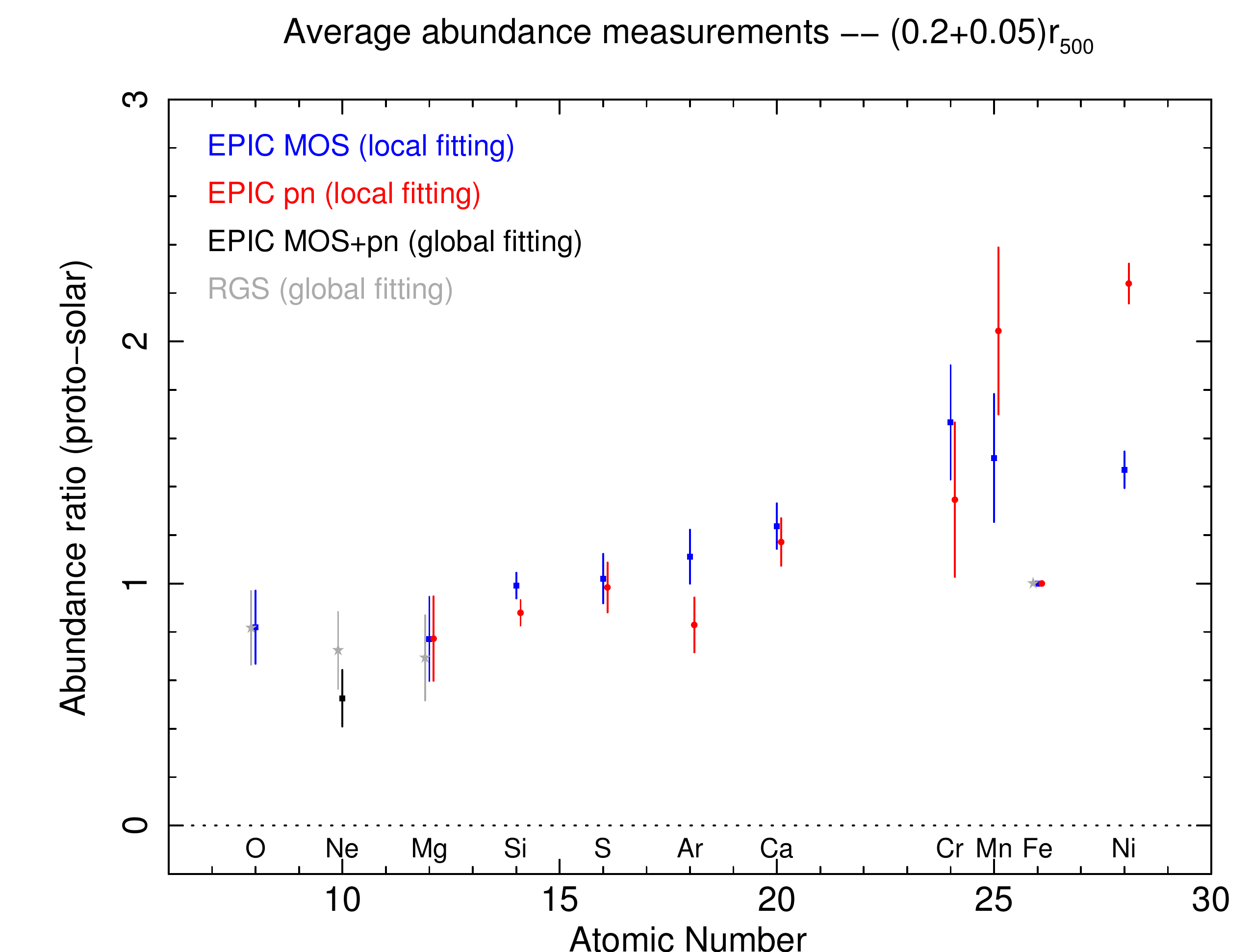}

        \caption{\textit{Left}: Average abundance ratios considering different (sub-)samples and/or different EPIC extraction regions. The O/Fe and Ne/Fe ratios are measured using RGS (cross-dispersion width of 0.8$'$). For these two ratios there is thus no distinction between the two hot samples or between the two full samples. The error bars incorporate the statistical errors ($\sigma_\text{stat}$) and the intrinsic scatters ($\sigma_\text{int}$ + 1$\sigma$). \textit{Right}: Average abundance ratios estimated from the $(0.05+0.2)r_{500}$ sample, measured independently by different combinations of instruments. The error bars incorporate the statistical errors ($\sigma_\text{stat}$), the intrinsic scatters ($\sigma_\text{int}$ + 1$\sigma$), and the uncertainties derived from the different EPIC extraction regions and/or subsamples ($\sigma_\text{region}$, see left panel).}
\label{fig:systematics}
\end{figure*}

Second,  we investigate whether the average abundance ratios change significantly  when considering different EPIC extraction regions and/or subsamples. The comparison of these ratios over the four (sub-)samples described in Sect. \ref{sect:select_abundances} is shown in Fig. \ref{fig:systematics} (left panel). The abundance ratios of all the elements are consistent, except for S/Fe and Ar/Fe, which we discuss more extensively in Sect. \ref{sect:elements}. For these two elements, we determine the systematic uncertainty $\sigma_\text{region}$ by artificially increasing their combined uncertainties $\sqrt{ \sigma_\text{stat}^2 + (\sigma_\text{int}\text{ + 1}\sigma)^2 + \sigma_\text{region}^2 }$, until they cover the discrepancies between the (sub-)samples, and make them all $\le$1$\sigma$ consistent.

Third, after correction for $\sigma_\text{int}$ and $\sigma_\text{region}$, we look for possible cross-calibration biases by comparing the average abundances estimated from the separate \textit{XMM-Newton} instruments (Fig. \ref{fig:systematics}, right panel). Three elements have MOS and pn abundance ratios that differ with more than 1$\sigma$ significance, and need an additional systematic uncertainty ($\sigma_\text{cross-cal}$, defined similarly to $\sigma_\text{region}$): Si/Fe, Ar/Fe, and Ni/Fe. The last two are the most striking: pn estimates the Ar/Fe and Ni/Fe ratios on average respectively $\sim$25\% lower and $\sim$52\% higher than MOS. A further discussion on the discrepancies found in these two ratios will be addressed in Sect. \ref{sect:elements}.

Fourth, since the conversion from the EW of a considered line to the abundance of its element strongly depends on the plasma temperature, a multi-temperature structure deviating from the \texttt{gdem} distribution may affect the abundance ratios. We investigate this dependency for the best EPIC observations of Perseus and M\,87 in Appendix \ref{sect:kT_models}.
Among the two continuous temperature distributions tested here \citep[which are thought to be the most reasonable to describe the thermal structure of the ICM; e.g.][]{2006A&A...452..397D}, we find that the deviations in the EPIC abundance ratios are marginal, well below the range of the other systematic uncertainties discussed above. Therefore, we do not consider this effect in the rest of this paper. \\

Assuming the systematic errors mentioned above to be roughly symmetric, we add them in quadrature to obtain the total uncertainties:
\begin{equation}
\sigma_\text{tot}^2 = \sigma_\text{stat}^2 + (\sigma_\text{int}\text{ + 1}\sigma)^2 + \sigma_\text{region}^2 + \sigma_\text{cross-cal}^2
.\end{equation}

Finally, we must note that further systematic uncertainties might still play a role. For example, we show in Appendix \ref{sect:RR_rates} that too simple approximations in the calculation of the emission processes might alter the line emissivities, and thus the abundances we measure. Therefore, we cannot exclude that future improvements in the currently used spectral fitting codes could still slightly affect the measurements we report here. Moreover, from some aspects \citep[e.g. \ion{Fe}{xvii} line ratios; see][]{2012A&A...539A..34D}, small deviations have been reported in the spectral modelling of CIE plasmas using either the SPEX code, or the APEC model (based on the AtomDB code). In terms of abundances, the discrepancies between the two codes may bring further uncertainties, at least for RGS measurements (de Plaa et al., to be submitted); however, APEC is a single-temperature model, which should be avoided in this kind of analysis (Sect. \ref{sect:spectral_analysis}). Moreover, this lack of multi-temperature distribution for APEC makes a direct comparison between the two codes difficult.

\begin{table}
\begin{centering}
\caption{Average abundance ratios estimated from the $(0.05+0.2)r_{500}$ sample, as well as their statistical, systematic, and total uncertainties. An absence of value ($-$) means that no further uncertainty was required (see text).}             
\label{table:systematics}
\resizebox{\hsize}{!}{
\begin{tabular}{c| c| c c c c |c}        
\hline \hline                
Element & Mean & $\sigma_\text{stat}$ & $\sigma_\text{int}$ & $\sigma_\text{region}$ & $\sigma_\text{cross-cal}$ & $\sigma_\text{tot}$ \\    
 & value &  &  &  &  &  \\    

\hline                        
O/Fe & $0.817$ & $0.018$ & $0.116 \pm 0.035$ & $-$ & $-$ & $0.152$ \\
Ne/Fe & $0.724$ & $0.028$ & $0.103 \pm 0.054$ & $-$ & $-$ & $0.159$ \\
Mg/Fe & $0.743$ & $0.010$ & $0.145 \pm 0.029$ & $-$ & $-$ & $0.174$ \\ 
Si/Fe & $0.871$ & $0.012$ & $0.031 \pm 0.022$ & $-$ & $0.018$ & $0.057$ \\
S/Fe & $0.984$ & $0.014$ & $0.042 \pm 0.026$ & $0.076$ & $-$ & $0.103$ \\ 
Ar/Fe & $0.88$ & $0.03$ & $-$ & $0.11$ & $0.09$ & $0.15$ \\
Ca/Fe & $1.218$ & $0.031$ & $(<0.091)$ & $-$ & $-$ & $0.096$ \\ 
Cr/Fe & $1.56$ & $0.19$ & $-$ & $-$ & $-$ & $0.19$ \\ 
Mn/Fe & $1.70$ & $0.22$ & $-$ & $-$ & $-$ & $0.22$ \\ 
Ni/Fe & $1.93$ & $0.12$ & $-$ & $-$ & $0.38$ & $0.40$ \\

\hline                                   
\end{tabular}}
\par\end{centering}
\end{table}

\section{Discussion}\label{sect:discussion}


In this work, we have derived the abundances in the cores of 44 galaxy clusters, groups, and ellipticals (CHEERS), using both the EPIC and RGS instruments. We have shown (Fig. \ref{fig:sample_kT_abun_others}) that the abundance ratios of O/Fe, Ne/Fe, Mg/Fe, Si/Fe, S/Fe, Ar/Fe, Ca/Fe, and Ni/Fe are quite uniform over the considered ranges of temperatures in the sample (0.6--8 keV). These results corroborate the study of \citet{2009A&A...508..565D}, who also found flat trends for Si/Fe and Ni/Fe independently of the considered clusters. This strongly suggests that regardless of their precise nature and of their different spatial scales, the physical processes that are responsible for the enrichment of the ICM must be the same for ellipticals, galaxy groups, and galaxy clusters. 

Unlike these ratios, the absolute Fe abundance is far from being uniform, and seems much more dependent on cluster history (Fig. \ref{fig:sample_kT_abun_Fe}). The scatter is more important in the inner regions ($0.05r_{500}$) of the core. 
A less scattered Fe abundance within $0.2r_{500}$ could suggest a flatter abundance distribution as we look away from the centre with a  similar level of enrichment outside the core of most objects. This is in agreement with the hypothesis of an early (pre-)enrichment, supported by recent \textit{Suzaku} observations of outskirts of clusters/groups \citep{2013Natur.502..656W,2015ApJ...811L..25S}.

The cool groups/ellipticals appear on average to be less Fe-rich than the hot clusters. In particular, it is interesting to note that nine hot clusters have an Fe abundance that is higher than proto-solar within $0.05r_{500}$, while, at the same scale, no cool group/elliptical has a similar feature (Fig. \ref{fig:sample_histo_Fe_best}). This trend has been already reported observationally \citep{2009MNRAS.399..239R} and in simulations \citep{2016MNRAS.456.4266L}. It might be explained by several scenarios:

\begin{itemize}
\item More massive objects are more efficient in retaining metals within their core (owing to their larger gravitational well or a less powerful AGN activity);
\item The more massive clusters are somewhat more efficient in injecting synthesised metals into the ICM;
\item The galaxies of the more massive clusters are somewhat more efficient in producing stars, and hence, SNe;
\item A more efficient cooling in group cores removes the enriched gas observed in X-ray.
\end{itemize}

While \citet{2016MNRAS.456.4266L} propose that the last scenario  explains the lack of metal-rich gas in lower-mass (hence, lower-temperature) objects, \citet{2009MNRAS.399..239R} explored the four possibilities, and argue that the the two first  are the most likely. In particular, the galactic outflows could be less efficient in releasing metals in the ICM of cooler groups or, alternatively, the AGN activity of the BCG could have helped to remove metals from their core \citep[see also][]{2016arXiv160304858Y}. However, we must emphasise that several of these mechanisms might co-exist, and that the list above is not necessarily exhaustive. For instance, \citet{2015ApJ...805....3E} recently found a hint of a positive correlation between the metallicity in low-mass clusters, and the morphological disturbance of their ICM (likely related to the dynamical activity of their galaxy members). 
Alternatively, cooler ICM might be more efficient in depleting ionic Fe (and probably other metals) into grains close to the brightest central galaxy, although \citet{2015MNRAS.447..417P} found hints of Fe depletion in the cores of more massive clusters as well. 

We must warn, however, that the large intrinsic scatters mentioned above prevent us from claiming any clear and significant trend on the absolute Fe abundances. Moreover, a more complicated thermal structure in groups/ellipticals than in more massive clusters cannot be excluded, and could lead to a slight but significant Fe bias, which would affect in priority the cooler objects in our sample.

We have also found an apparent variation in the Fe abundance and temperature gradients (i.e. between $0.05r_{500}$ and $0.2r_{500}$) in the hot clusters. These differences could be related to the individual cluster enrichment histories or to other parameters, such as the cooling rate. Linking the history of each cluster/group to its Fe budget requires a more careful spatial study of the Fe distribution. Establishing radial profiles for the entire sample is beyond the scope of this work, but will be addressed in a separate paper.

From the stacked results of our $(0.05+0.2)r_{500}$ sample, we have estimated the average abundance ratios and their respective total uncertainties (statistical and systematic). This also includes Cr/Fe and Mn/Fe, which we have detected within >4$\sigma$ significance with MOS and pn independently. To our knowledge, this is the first time that Mn has been firmly detected in the ICM. For comparison, \citet{2006A&A...449..475W} already detected Cr in 2A\,0335+096 within 2$\sigma$ (but were unable to detect Mn), while Cr and Mn have been detected in Perseus within 5$\sigma$ and 1$\sigma$, respectively \citep{2009ApJ...705L..62T}. It is also striking to note that we do not see any emission line feature around $\sim$3.5 keV in the stacked EPIC spectra (Fig. \ref{fig:residuals} bottom, Fig\ref{fig:residuals_hotcold}) contrary to several claims from recent studies, in particular \citet{2014ApJ...789...13B} and \citet{2014PhRvL.113y1301B}, whose total EPIC net exposure times are $\sim$3 Ms and $\sim$1.5 Ms, respectively (i.e. less than in this work). Such an apparent non-detection is thus very interesting to report, since it might challenge the hypothesis of decaying sterile neutrinos, known as a dark matter candidate,  being observed in the ICM. We note that the dark matter interpretation is far from being the only possible explanation of an emission line at $\sim$3.5 keV \citep[e.g.][]{2015A&A...584L..11G}, and our non-detection could  also be explored in the context of these other possibilities. However, this question is not the initial purpose of this present study, and a more detailed investigation of our stacked spectra around $\sim$3.5 keV, and consequent discussions, are left to a future paper.

\subsection{Discrepancies in the S/Fe, Ar/Fe and Ni/Fe ratios}\label{sect:elements}

The average S/Fe ratio shows a slight but significant enhancement in the cool objects within $0.05r_{500}$ compared to the hot objects within $0.2r_{500}$ (Fig. \ref{fig:systematics}). From Fig. \ref{fig:sample_kT_abun_others} (left, third  panel), it is   clearly  that M\,49 ($kT_\text{mean} \sim 1.148$ keV) and A\,3581 ($kT_\text{mean} \sim 1.637$ keV) largely contribute to this higher S/Fe ratio measured in the cool subsample because their statistical errors are small compared to the other cool objects. Moreover, the >1$\sigma$ discrepancy between the cool and hot measured S/Fe ratios disappears when considering the same radius ($0.05r_{500}$) for all the objects.

The Ar/Fe discrepancy observed in Fig. \ref{fig:systematics} (left) is more intriguing, since a larger aperture seems to lower its measurement. This trend is difficult to interpret. A change in the relative Ar to Fe radial distribution in the ICM cannot be excluded, although we would then expect it for other elements as well. A full study of the abundance radial profiles of the sample will be performed in a future paper. Furthermore, it also appears from Fig. \ref{fig:systematics} (right) that MOS and pn measure significantly different Ar/Fe values, even after taking account of the uncertainty described above (i.e. $\sigma_\text{region}$). The reason for this second Ar/Fe discrepancy is again challenging to clearly identify, but it is very likely due to imperfections in the calibration of the EPIC instruments.

As seen in Fig. \ref{fig:systematics} (right), the large MOS-pn discrepancy in the Ni/Fe abundance ratio prevents us from deriving a precise measurement. This discrepancy is worrying, but can be explained by imperfections in the cross-calibration of the two instruments. Alternatively, and perhaps more likely, the high energy band around the Ni-K transitions is significanly affected by the instrumental background (as the flux of the cluster emission sharply decreases at high energies). This hard particle background (already mentioned in Sect. \ref{subsect:bkg_modelling}) has a different spectral shape in MOS and pn, which might even vary with time, thus between observations. In particular, an instrumental line (Cu K$\alpha$) is known to affect pn at a rest-frame energy of $\sim$8 keV \citep{2015A&A...575A..37M}. Despite our efforts to carefully estimate the background, that line might interfere with the Ni-K line in several observations, making a proper modelling of the Ni-K line impossible, and hence, boosting the Ni absolute abundance in pn. In this context, it can be instructive to compare our Ni/Fe measurements with those of \textit{Suzaku}, which has a lower relative hard particle background. \citet{2007PASJ...59..299S} (A\,1060) and \citet{2009ApJ...705L..62T} (Perseus) reported ratios of $\sim$1.3$\pm$0.4 and $\sim$1.11$\pm$0.19, respectively (after rescaling to the proto-solar values). Although these measurements might be also be affected by further uncertainties (e.g. the choice of the spectral modelling, Sect. \ref{sect:systematics}), they appear to be consistent with the Ni/Fe average ratio measured with MOS is this work, favouring our above supposition that MOS is more trustworthy than pn for measuring Ni/Fe. However, in order to be conservative, we prefer to retain the pn value as a possible result and, therefore, we keep large systematic uncertainties for Ni/Fe. We finally note  that, unsurprisingly, Ni/Fe cannot be constrained in the cool objects (Fig. \ref{fig:systematics}, left) because the gas temperature is too low to excite Ni-K transitions.

\subsection{Comparison with the proto-solar abundance ratios}\label{sect:proto-solar}

\begin{figure}[!]
        \centering
                \includegraphics[width=0.49\textwidth]{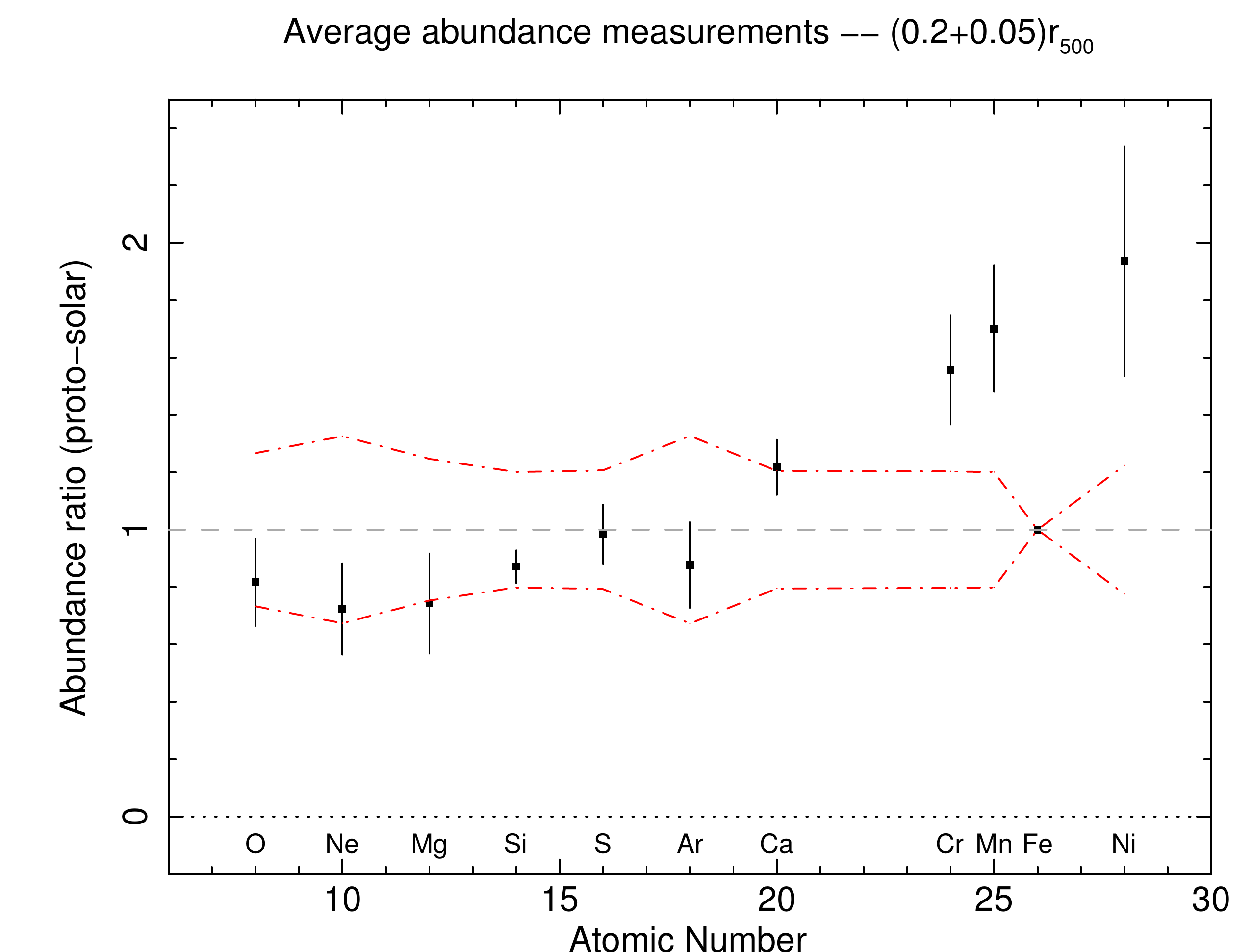}

        \caption{Average abundance ratios in our $(0.05+0.2)r_{500}$ sample (black squares) versus proto-solar abundances (grey dashed line) and their 1$\sigma$ uncertainties \citep[red dash-dotted lines, adapted from][]{2009LanB...4B...44L}. }
\label{fig:cluster_vs_solar}
\end{figure}

In Fig. \ref{fig:cluster_vs_solar} (black squares), we report our final X/Fe abundance pattern measured in the $(0.05+0.2)r_{500}$ sample, accounting for all the systematic uncertainties discussed earlier in this paper. At first glance, most of the abundance ratios measured in the ICM look significantly different from the proto-solar abundance ratios. Indeed, if we fit a constant to our abundance pattern (dashed grey line), we obtain a $\chi^2$ of 43.1 for 10 degrees of freedom, in poor agreement with the abundance ratios in the ICM. However, as shown by the red dash-dotted lines \citep[adapted from][]{2009LanB...4B...44L}, the solar abundance ratios also suffer from large uncertainties, typically about 20--25\%. When comparing the two sets of abundance ratios taken with their respective uncertainties, we find that the O/Fe, Ne/Fe, Mg/Fe, Si/Fe S/Fe, Ar/Fe, and Ca/Fe ratios measured in the ICM are consistent within 1$\sigma$ with the proto-solar values ($\pm$1$\sigma$). The Cr/Fe  abundance ratios measured in the ICM are consistent  within 2$\sigma$ with the proto-solar
values ($\pm$1$\sigma$), and the Mn/Fe and
Ni/Fe abundance ratios are consistent within 3$\sigma$.  

Given these considerations, whether the chemical enrichment in the ICM is similar to the chemical enrichment of the solar neighbourhood is not a trivial question to solve. As mentioned above, while most of the relative abundances appear to be consistent with being proto-solar, Ni/Fe, Mn/Fe, and perhaps Cr/Fe seem to be significantly enhanced. This result might be of interest since significant different abundance ratios in the ICM means that the fraction of SNIa (or conversely SNcc) responsible for the ICM enrichment differs from that of the Galactic enrichment. We will address this discussion  in greater detail in Paper II. We recall, however, that the abundances of interest in this context (Cr, Mn, and Ni) are not well constrained in X-ray, given the current instrument capabilities. Moreover, as specified in Sect. \ref{sect:systematics},  we do not exclude that differences in current atomic codes might bring further systematic uncertainties to the measurements reported in this work. Therefore, whether our abundance ratios are significantly more accurate than the proto-solar estimates should still be considered  an open question.

\subsection{Current limitations and future prospects}\label{sect:future}

As we have shown throughout this paper, our excellent data quality ($\sim$4.5 Ms and $\sim$3.7 Ms of total net exposure time for EPIC MOS and pn, respectively) provides very accurate abundance measurements in the ICM of cool-core galaxy clusters and groups. Therefore, these data should be a legacy for any future work directly or indirectly related to the chemical enrichment in the ICM. However, our study clearly reveals that the abundance ratios of some elements of interest are still poorly constrained. In particular, the Cr/Fe, Mn/Fe, and Ni/Fe ratios in our study appear higher than  the solar neighbourhood and are thus crucial to study in detail, which is currently challenging given the limited spectral resolution of CCDs and the large cross-calibration uncertainties (at least for Ni/Fe) that we emphasise here.

In this work we probably reach the instrumental limitations of \textit{XMM-Newton} in terms of abundance determination, and thus stacking more data will have very little impact on the current accuracy of our already existing measurements. Indeed, in our sample, the statistical uncertainties are already marginal compared to the systematic ones. In addition to further efforts in calibrating the instruments and improving the atomic databases, it is clear that a sensibly higher X-ray spectral resolution is now needed.

Such an improvement can be reached with micro-calorimeter spectrometers, which should be on board the next generation of X-ray observatories. In particular, the Japanese X-ray observatory \textit{Hitomi} \citep[formerly named \textit{ASTRO-H};][]{2014arXiv1412.2351T} has been able to resolve, for instance, K-shell and also L-shell Ni lines with a limited instrumental background, and should thus reduce the Ni/Fe uncertainties to a few per cent. Unfortunately, owing to a loss of contact a few weeks after launch, the fate of the mission is now unclear. Alternatively, the X-IFU instrument, which will be on board the \textit{Athena} observatory \citep{2013arXiv1306.2307N}, will greatly improve the spectral resolution currently achieved with \textit{XMM-Newton} to $\sim$2.5 eV. Undoubtedly, this upcoming mission will allow a significant step forward in such an analysis, especially if improvements in atomic data are also carried out.

\section{Conclusions}\label{sect:conclusion}

In this paper, we have used the \textit{XMM-Newton} EPIC and RGS instruments to investigate the Fe abundance and abundance ratios of O/Fe, Ne/Fe, Mg/Fe, Si/Fe, S/Fe, Ar/Fe, Ca/Fe, Cr/Fe, Mn/Fe, and Ni/Fe in the central regions of 44 cool-core galaxy clusters, groups, and elliptical galaxies (CHEERS). Our main results can be summarised as follows.

\begin{itemize}

\item The X/Fe abundance ratios appear quite uniform over the mean temperature range of our sample. This confirms previous results, and indicates that no matter what the physical mechanisms responsible for the enrichment are, they must be very similar in enriching the ICM of ellipticals, groups, and clusters of galaxies.

\item By stacking all the EPIC spectra of our sample (within $0.2r_{500}$ when possible, within $0.05r_{500}$ otherwise), we were able to derive abundances of Cr/Fe and Mn/Fe independently with MOS and pn, with >4$\sigma$ significance. While Cr had been already detected in the past, this is the first time that a firm detection of Mn in the hot ICM has been reported. 

\item Contrary to recent claims, and despite the large net exposure time ($\sim$4.5 Ms) of our combined data, we do not see any emission line at $\sim$3.5 keV. Although a deeper investigation will be addressed in a future paper, this might challenge the possibility of decaying sterile neutrinos, a dark matter candidate,  being observed in the ICM.

\item The Fe abundance varies between 0.2--2 times the proto-solar values, and shows an important scatter, especially within a radius of $0.05r_{500}$ ($\sim$30--40\%). Looking at smaller ($0.05r_{500}$) and larger ($0.2r_{500}$) central regions in a subsample of hot clusters, it appears that the Fe peak sharpens and the temperature drop flattens as the mean cluster temperature decreases. Clearly, these various Fe abundances must depend on individual clusters histories, and complete abundance radial profiles will be investigated in greater detail in a future work.

\item Having benefited from a large total net exposure time ($\sim$4.5 Ms) and having processed a very careful estimation of the systematic effects that could affect our measurements, we have shown that the systematic uncertainties clearly dominate over the statistical ones. Taking these systematic uncertainties into account, most of the ICM abundance ratios measured in this work are consistent with the proto-solar abundance ratios. Notable exceptions are Mn/Fe, Ni/Fe, and perhaps Cr/Fe, which are found to be significanly higher in the ICM than in the solar neighbourhood.

\item Overall, our careful analysis demonstrates that stacking more observations would not further improve the accuracy of our results, and, more generally, that we have probably reached the limits of the current X-ray capabilities (in particular \textit{XMM-Newton}) for this science case. Therefore, our data constitute the most accurate abundance ratios ever measured in the ICM, and should be a legacy for future work. Using the results presented in this paper, a full discussion on the role of the SNIa and SNcc in the context of both the proto-solar and the ICM enrichments will be addressed in Paper II. However, a more accurate comparison between the local Galactic enrichment and the ICM enrichment in the local Universe will require improvements in atomic data, as well as  better calibration of the instruments. In parallel to these needs for improvements, the upcoming X-ray observatories should further improve the accuracy of the abundance measurements, and thus help to solve the puzzle of the chemical enrichment in the hot ICM.

\end{itemize}

\begin{acknowledgements}
The authors would like to thank the referee Marten van Kerkwijk for his helpful comments and suggestions. This work is partly based on the \textit{XMM-Newton} AO-12 proposal ``\emph{The XMM-Newton view of chemical enrichment in bright galaxy clusters and groups}'' (PI: de Plaa), and is a part of the CHEERS (CHEmical Evolution Rgs cluster Sample) collaboration. The authors thank its members, as well as Liyi Gu and Craig Sarazin for helpful discussions. P.K. thanks Steve Allen and Ondrej Urban for support and hospitality at Stanford University. Y.Y.Z. acknowledges support by the German BMWI through the Verbundforschung under grant 50\,OR\,1506. This work is based on observations obtained with \textit{XMM-Newton}, an ESA science mission with instruments and contributions directly funded by ESA member states and the USA (NASA). The SRON Netherlands Institute for Space Research is supported financially by NWO, the Netherlands Organisation for Space Research.
\end{acknowledgements}

\bibliography{Core_abundances-I_hk}{}
\bibliographystyle{aa}




\newpage

\appendix

\section{EPIC absorption column densities}\label{sect:N_H}

\begin{figure}[!]
        \centering
                \includegraphics[width=0.49\textwidth]{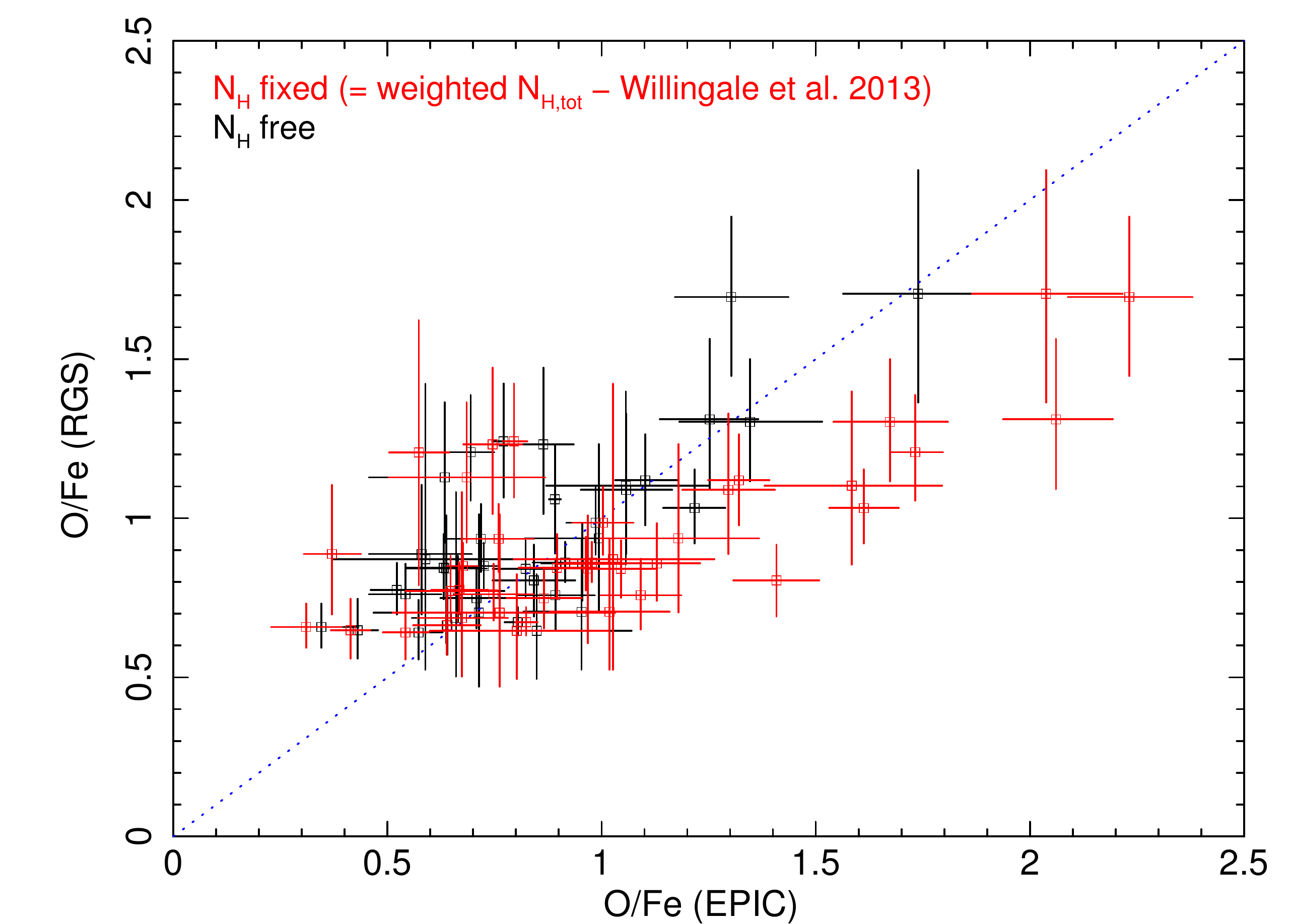}

        \caption{Comparison between the EPIC and RGS measurements of the O/Fe abundance ratio in most objects of our $(0.05+0.2)r_{500}$ sample. The blue dotted line shows the one-to-one EPIC-RGS correspondence. In our fits we alternatively fix the $N_\text{H}$ to the weighted neutral+molecular values calculated from \citet{2013MNRAS.431..394W}, and leave it free within the ranges given by Eq. (\ref{eq:N_H}). The two approaches are shown in red and black, respectively.}
\label{fig:N_H}
\end{figure}

In the \texttt{hot} model used in this work to mimic absorption of X-rays through interstellar material (Sect. \ref{sect:spectral_analysis}), we initially fixed the hydrogen column density $N_\text{H}$ to the weighted value $N_\text{H,tot}$ of both the neutral \citep[$N_\ion{H}{i}$,][]{2005A&A...440..775K} and the molecular \citep[$N_{\text{H}_2}$,][]{1998ApJ...500..525S} materials, calculated using the method of \citet{2013MNRAS.431..394W}\footnote{http://www.swift.ac.uk/analysis/nhtot/index.php}. However, this approach often gives poor fits in the soft band of our EPIC spectral modelling, by significantly under- or overestimating the flux of its continuum. In order to compensate this  effect, the O abundance is often biased consequently by the fits. Fig. \ref{fig:N_H} (red data points) clearly illustrates that some objects have their EPIC O/Fe ratio significantly offset from the corresponding  RGS values.

The EPIC-RGS correlation for the O/Fe ratio is clearly improved if we free the $N_\text{H}$ (Fig. \ref{fig:N_H}, black data points). Similarly, most of the fits are improved in terms of C-stat/d.o.f. However, keeping $N_\text{H}$ as a free parameter without any further constraint is quite dangerous, and might lead to unphysical results. In order to remain reasonably consistent with the estimated values of $N_\ion{H}{i}$ and $N_\text{H,tot}$ mentioned above, we allow $N_\text{H}$ to take values within the following arbitrary limits: 
\begin{equation}\label{eq:N_H}
N_\ion{H}{i} - 5\times 10^{19} \text{ cm}^{-2} \le N_\text{H} \le N_\text{H,tot} + 1 \times 10^{20} \text{ cm}^{-2}.
\end{equation}

These upper and lower ranges allow limited deviations also around  $N_\ion{H}{i}$ and $N_\text{H,tot}$. Since constraining a free parameter within a narrow range can lead to problems in evaluating the statistical errors, we perform a grid search of fixed  $N_\text{H}$ values (taken within the limits mentioned above), and select the one that gives the lowest C-stat/d.o.f. to the fits. Despite all these precautions,  it should also be kept in mind that the O abundance measured in clusters with EPIC is also affected by the oxygen absorption in the interstellar medium, which in turn depends on $N_\text{H}$ \citep[e.g.][]{2004A&A...423...49D}. Similarly, the measured O abundance in the ICM may be also affected by the foreground thermal X-ray emission.

\section{Radiative recombination corrections}\label{sect:RR_rates}

The version of SPEX that is used in this work calculates the line emissivities assuming that the radiative recombination (RR) rates of the cluster emission can be expressed as a power law of the electron temperature \citep{1981A&AS...45...11M,1985A&AS...62..197M}.
However, this approximation has turned out to be too simplified at high temperature. A more accurate calculation of the RR rate coefficients has been done by \citet{2006ApJS..167..334B}, and parametrised by \citet{2016A&A...587A..84M} as a function of the temperature $T$ in the form
\begin{equation}
R(T) \propto  T^{-b_0 -c_0 \ln T} \left( \frac{1+a_2 T^{-b_2}}{1+a_1 T^{-b_1}} \right),
\end{equation}
where $a_0$, $b_0$, $c_0$, $a_1$, $a_2$, $b_1$, and $b_2$ are constant (fitted) parameters.

\begin{figure}[!]
        \centering
                \includegraphics[width=0.49\textwidth]{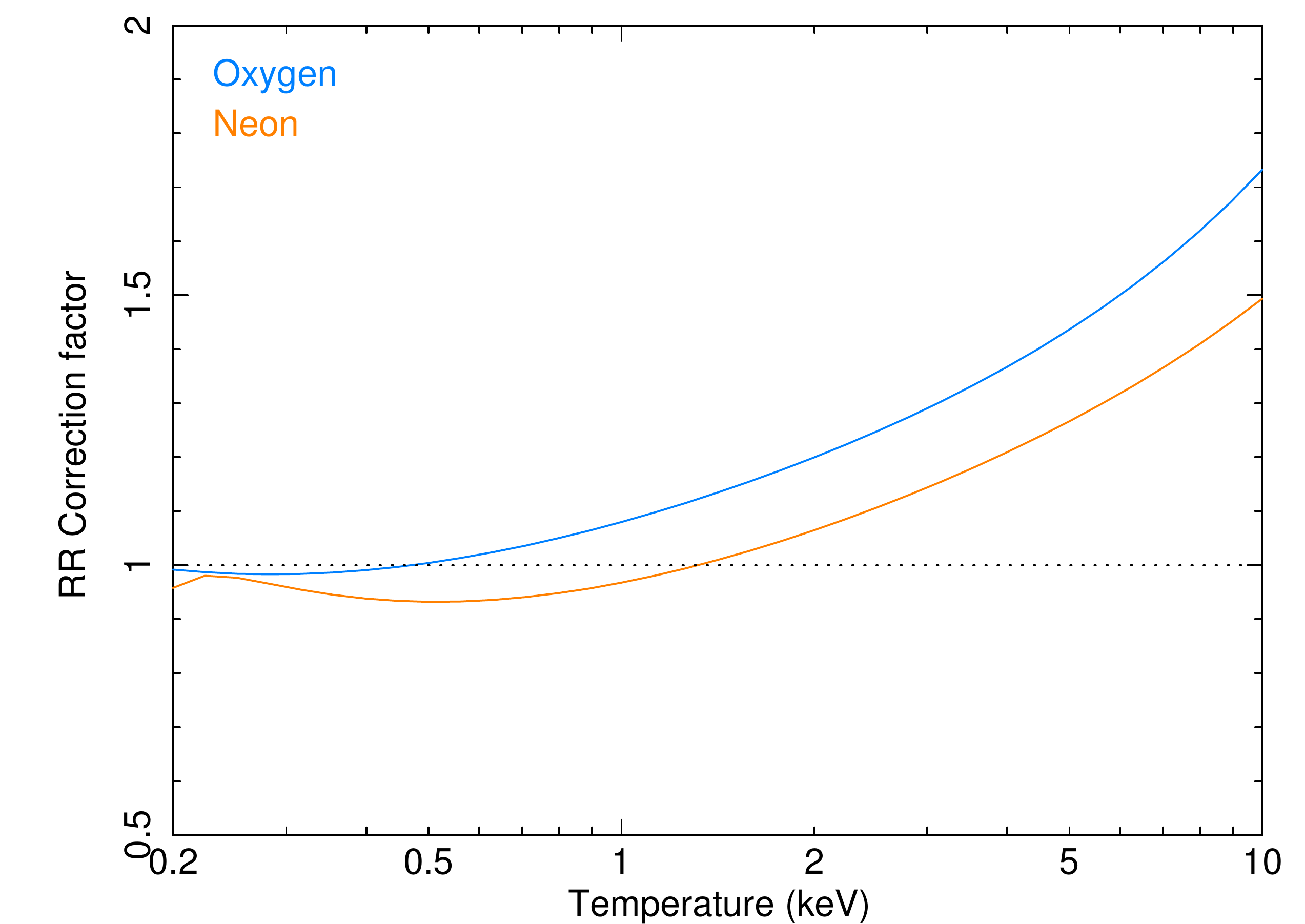}

        \caption{Calculated radiative recombination correction factors of H-like Lyman $\alpha$ lines of O and Ne as a function of the cluster (mean) temperature \citep[adapted from the results of][]{2016A&A...587A..84M}.}
\label{fig:RR_rates}
\end{figure}

Since the RR rates directly affect the line emissivities, which in turn affect our estimated abundances, the RR updated model of \citet{2016A&A...587A..84M} must be taken into account in our analysis, even though its implementation into SPEX is yet to come. Knowing that the O and Ne emission lines seen in clusters spectra are dominated by H-like Lyman $\alpha$ transitions, and that these two elements are the most affected by changing RR rates, we correct their abundances by computing the change in flux of their H-like Lyman $\alpha$ lines from the old RR calculations (i.e. used in the current SPEX version) to the new ones. This RR correction factor is shown (again, for O and Ne) in Fig. \ref{fig:RR_rates} as a function of the plasma temperature, and is to be multiplied by the measured O and Ne abundances of each object in our sample.

Figure \ref{fig:RR_rates} clearly shows that better calculations of the RR rates can lead to significant increases of the estimated O and Ne abundances in hot clusters. After applying this RR correction factor for each source, we find that, on average, the O/Fe and Ne/Fe abundance ratios increase by $\sim$20\% and $\sim$9\%, respectively. We note, however, that reprocessing the whole analysis presented in this paper by using the uncorrected O and Ne abundance values does not affect our main conclusions, since we keep large systematic uncertainties in the final abundance ratios.

\section{Effects of the temperature distribution on the abundance ratios}\label{sect:kT_models}

\begin{figure*}[!]
        \centering
                \includegraphics[width=0.49\textwidth]{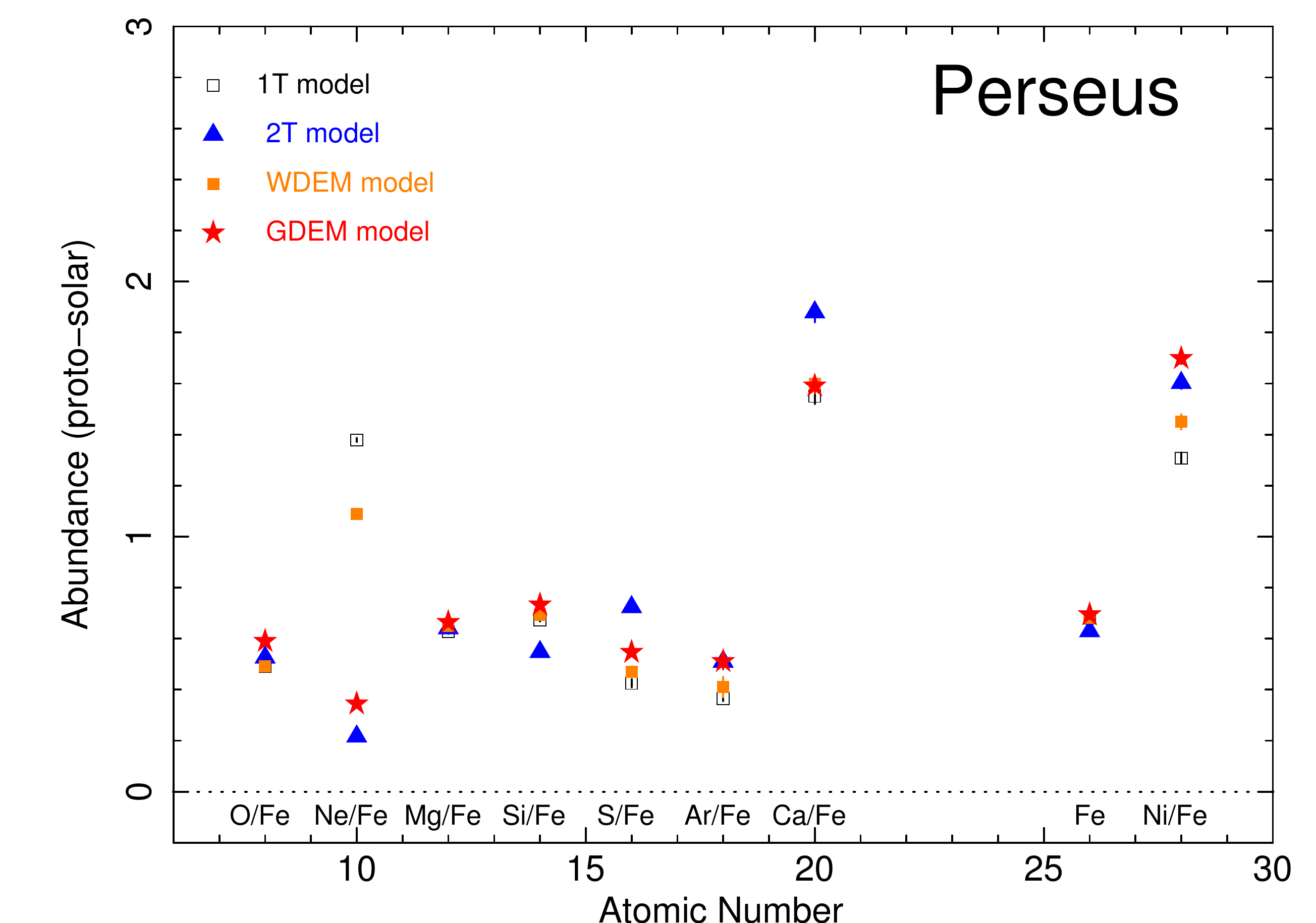}
                \includegraphics[width=0.49\textwidth]{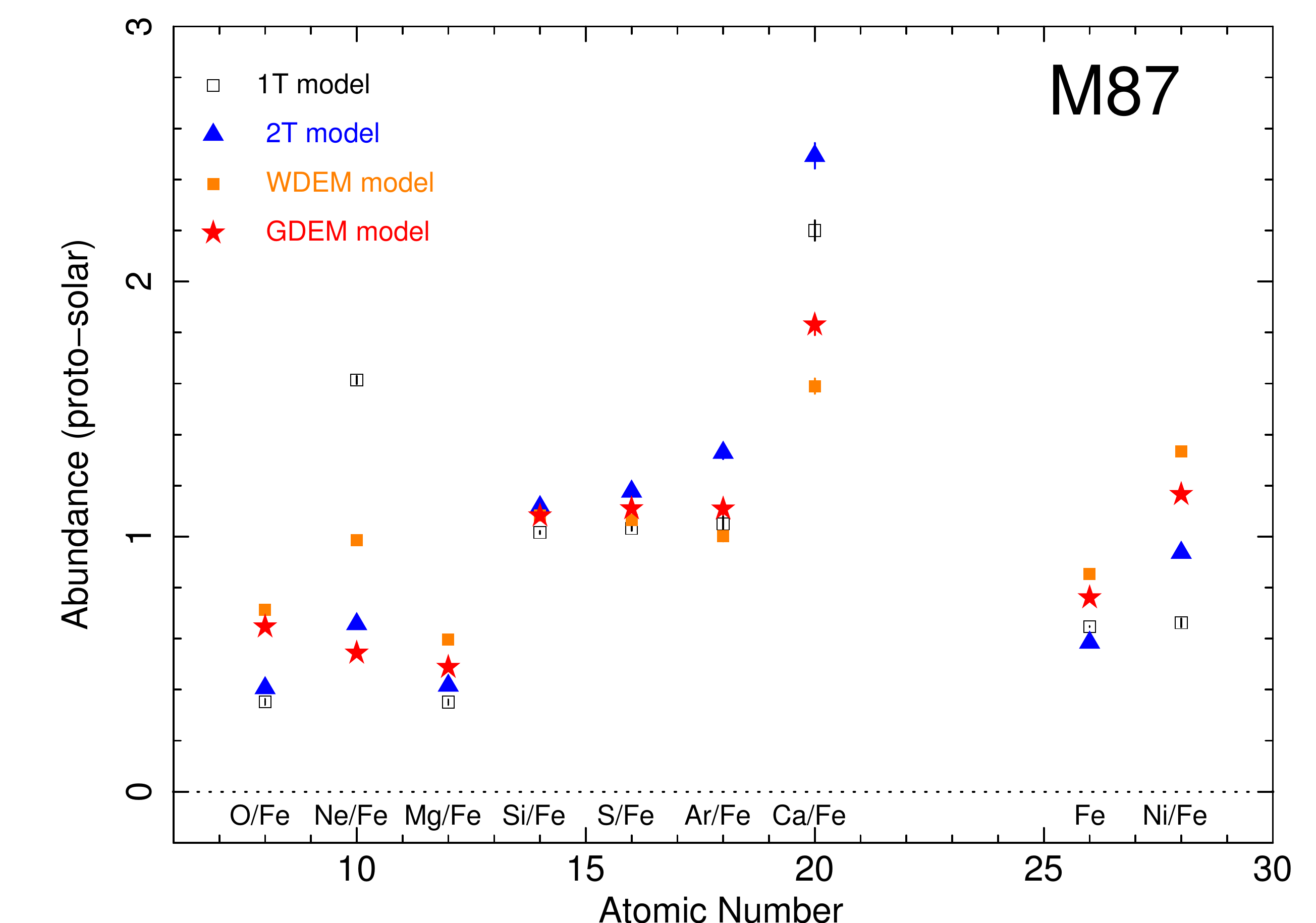}

        \caption{EPIC abundance measurements in the two best-quality observations of our sample, based on global fittings (see also Table \ref{table:kT_models}). Four thermal models are successively considered: 1T (black empty squares), 2T (blue filled squares), \texttt{wdem} (orange filled squares), and \texttt{gdem} (red filled squares). The Fe abundance is given in absolute values, while the other abundances are given relative to Fe. \textit{Left}: Perseus (0.2$r_{500}$). \textit{Right}: M\,87 (0.05$r_{500}$).}
\label{fig:kT_models}
\end{figure*}

As already specified in Sect. \ref{sect:spectral_analysis}, the measured absolute abundances (in particular Fe) are in principle sensitive to the choice of the  thermal model used in the fits (single- vs. multi-temperature).  Among multi-temperature models, the assumed temperature distribution might also affect the measured (X/Fe) abundances. We explore this possibility by successively fitting the best-quality observations of Perseus and M\,87 (which both have excellent statistics but rather different mean temperatures)  with a 1T (i.e. \texttt{cie}), a 2T (i.e. \texttt{cie+cie}), and a power-law differential emission measure model\footnote{This model is thought to reproduce quite well the temperature structure in the ICM of most cool-core objects, but has not been used in this work owing to computing time.} \citep[\texttt{wdem}; see e.g.][]{2004A&A...413..415K}. The results are shown in Fig. \ref{fig:kT_models} and Table \ref{table:kT_models}. From Fig. \ref{fig:kT_models}, it clearly appears that the abundance pattern depends on the considered thermal model. In particular, Ne/Fe varies a lot (i.e. by more than a factor of 6 for Perseus and by almost a factor of 3 for M\,87) because the Ne abundance parameter from the models is used by the fits to compensate the EPIC residuals in the Fe-L complex \citep[e.g.][]{2006A&A...452..397D}. This illustrates that the EPIC estimate of Ne/Fe cannot be interpreted as a reliable Ne abundance (Sect. \ref{sect:select_abundances}). Striking differences in the Ca/Fe and Ni/Fe ratios considering the four different models should also be noted; for instance, a considerably high Ca/Fe ratio is measured by the 1T and/or 2T model(s).

\begin{table*}
\begin{centering}
\caption{Comparison of the abundance results obtained by performing global fittings to the best-quality EPIC spectra of Perseus (0.2$r_{500}$) and M\,87 (0.05$r_{500}$). Four different temperature distributions of the CIE model are successively considered (1T, 2T, \texttt{wdem}, and \texttt{gdem}; see also Fig. \ref{fig:kT_models}).}             
\label{table:kT_models}
 \setlength{\tabcolsep}{15pt}
\begin{tabular}{c | c c c c}        
\hline \hline                
Element & 1T & 2T & \texttt{wdem} & \texttt{gdem}  \\    
\hline
 & \multicolumn{4}{c}{Perseus}  \\    

\hline                        
O/Fe & $0.491 \pm 0.006$ & $0.526 \pm 0.007$ & $0.491 \pm 0.007$ & $0.589 \pm 0.006$ \\
Ne/Fe & $1.379 \pm 0.008$ & $0.217 \pm 0.012$ & $1.090 \pm 0.009$ & $0.344 \pm 0.008$ \\
Mg/Fe & $0.628 \pm 0.010$ & $0.641 \pm 0.012$ & $0.651 \pm 0.010$ & $0.663 \pm 0.010$ \\
Si/Fe & $0.673 \pm 0.006$ & $0.548 \pm 0.011$ & $0.693 \pm 0.006$ & $0.732 \pm 0.005$ \\
S/Fe & $0.425 \pm 0.016$ & $0.724 \pm 0.005$ & $0.470 \pm 0.010$ & $0.546 \pm 0.008$ \\
Ar/Fe & $0.365 \pm 0.010$ & $0.51 \pm 0.03$ & $0.41 \pm 0.04$ & $0.51 \pm 0.03$ \\
Ca/Fe & $1.55 \pm 0.03$ & $1.88 \pm 0.04$ & $1.60 \pm 0.03$ & $1.59 \pm 0.03$ \\
Fe & $0.6802 \pm 0.0011$ & $0.6296 \pm 0.0012$ & $0.6808 \pm 0.0015$ & $0.6934 \pm 0.0012$ \\
Ni/Fe & $1.308 \pm 0.022$ & $1.603 \pm 0.024$ & $1.45 \pm 0.03$ & $1.698 \pm 0.022$ \\
C-stat / d.o.f. & 39\,168 / 1963 & 25\,669 / 1961 & 25\,124 / 1962 & 32\,481 / 1962 \\

\hline
 & \multicolumn{4}{c}{M\,87} \\    

\hline                        
O/Fe & $0.352 \pm 0.012$ & $0.406 \pm 0.007$ & $0.713 \pm 0.008$ & $0.646 \pm 0.009$ \\
Ne/Fe & $1.614 \pm 0.016$ & $0.657 \pm 0.015$ & $0.987 \pm 0.013$ & $0.543 \pm 0.011$ \\
Mg/Fe & $0.351 \pm 0.011$ & $0.416 \pm 0.010$ & $0.597 \pm 0.009$ & $0.487 \pm 0.009$ \\
Si/Fe & $1.016 \pm 0.006$ & $1.116 \pm 0.006$ & $1.084 \pm 0.006$ & $1.081 \pm 0.007$ \\
S/Fe & $1.032 \pm 0.009$ & $1.177 \pm 0.009$ & $1.065 \pm 0.008$ & $1.109 \pm 0.009$ \\
Ar/Fe & $1.05 \pm 0.03$ & $1.329 \pm 0.025$ & $1.002 \pm 0.021$ & $1.108 \pm 0.023$ \\
Ca/Fe & $2.20 \pm 0.04$ & $2.493 \pm 0.05$ & $1.59 \pm 0.03$ & $1.83 \pm 0.04$ \\
Fe & $0.6472 \pm 0.0020$ & $0.5847 \pm 0.0016$ & $0.8532 \pm 0.0020$ & $0.7601 \pm 0.0017$ \\
Ni/Fe & $0.662 \pm 0.017$ & $0.937 \pm 0.014$ & $1.334 \pm 0.015$ & $1.165 \pm 0.016$ \\
C-stat / d.o.f. & 19\,034 / 914 & 13\,602 / 912 & 9\,474 / 913 & 8\,680 / 913 \\

\hline                                   
\end{tabular}
\par\end{centering}
\end{table*}

Despite these considerations, and the fact that the real temperature distributions in the ICM is unknown, considering a continuous distribution is clearly more realistic than only one or two unique temperatures. By comparing only the \texttt{gdem} and \texttt{wdem} models, we note that they give very similar abundance ratios for both Perseus and M\,87. Except for Ne/Fe, the largest difference is found for Ni/Fe, and is clearly smaller than the range of systematic uncertainties affecting the measurements. Therefore, using a \texttt{wdem} model instead of a \texttt{gdem} model should have a limited impact on our EPIC final results. For comparison, de Plaa et al. (to be submitted) find that such an effect on the O/Fe ratio derived from RGS is always smaller than 20\%. Therefore, considering further uncertainties related to the thermal models is not necessary for the purpose of this work.

\section{Best-fit temperature and abundances}\label{sect:full_results}

In Table \ref{table:full_results} we present the full results of our best-fit parameters ($kT$, $\sigma_T$, the absolute Fe abundance, and the abundance ratios of O/Fe, Ne/Fe, Mg/Fe, Si/Fe, S/Fe, Ar/Fe, Ca/Fe, and Ni/Fe) for each object of  CHEERS, within a radius of $0.05r_{500}$. When possible (hot clusters), we indicate the parameters extracted from $0.2r_{500}$ as well. The O/Fe and Ne/Fe abundances have been corrected from updated RR calculations, as described in Appendix \ref{sect:RR_rates}.

\onecolumn

\fontsize{8}{8}
\begin{longtab}
\begin{landscape}
\begin{longtable}{llccccccccccc}
\caption{\label{table:full_results}Summary of the best-fit temperatures, $\sigma_T$ and abundances for all the clusters/groups/ellipticals of the CHEERS. The $kT$, $\sigma_T$ parameters and the Fe absolute abundance are measured with EPIC, based on global fittings, as described in Sect. \ref{sect:global_fits}. The Mg/Fe, Si/Fe, S/Fe, Ar/Fe, Ca/Fe and Ni/Fe abundance ratios as well as their error bars are measured using EPIC, based on local fittings, as described in Sect. \ref{sect:local_fits}. The O/Fe and Ne/Fe abundance ratios are measured using RGS, and do not depend on the considered EPIC extraction region ($=$). When parameters are unrealistic or fully uncertain, we do not report any measurement ($-$).}\\
\hline\hline
Source & Region & $kT$ & $\sigma_T$ & O/Fe & Ne/Fe & Mg/Fe & Si/Fe & S/Fe & Ar/Fe & Ca/Fe & Fe & Ni/Fe\\
  & ($r_{500}$) & (keV) &  &   &   &   &   &   &   &   &   &  \\
\hline
\endfirsthead
\caption{continued.}\\
\hline\hline
Source & Region & $kT$ & $\sigma_T$ & O/Fe & Ne/Fe & Mg/Fe & Si/Fe & S/Fe & Ar/Fe & Ca/Fe & Fe & Ni/Fe\\
  & ($r_{500}$) & (keV) &  &   &   &   &   &   &   &   &   &  \\
\hline
\endhead
\hline
\endfoot
2A 0335 & 0.2 & $2.661 \pm 0.004$ & $0.2433 \pm 0.0013$ & $0.88 \pm 0.09$ & $0.97 \pm 0.14$ & $0.69 \pm 0.15$ & $0.82 \pm 0.10$ & $0.83 \pm 0.14$ & $0.87 \pm 0.27$ & $1.04 \pm 0.25$ & $0.765 \pm 0.004$ & $2.0 \pm 1.0$ \\
                         & 0.05 & $2.265 \pm 0.005$ & $0.2116 \pm 0.0014$ & $=$ & $=$ & $0.72 \pm 0.20$ & $0.83 \pm 0.10$ & $0.96 \pm 0.15$ & $1.01 \pm 0.26$ & $1.1 \pm 0.3$ & $0.882 \pm 0.007$ & $3.4 \pm 1.6$ \\

A 85 & 0.2 & $5.715 \pm 0.018$ & $0.354 \pm 0.003$ & $0.80 \pm 0.11$ & $0.80 \pm 0.16$ & $0.76 \pm 0.06$ & $0.76 \pm 0.18$ & $0.92 \pm 0.06$ & $0.98 \pm 0.13$ & $1.0 \pm 0.6$ & $0.733 \pm 0.006$ & $1.7 \pm 1.2$ \\
        & 0.05 & $4.432 \pm 0.020$ & $0.300 \pm 0.004$ & $=$ & $=$ & $0.74 \pm 0.07$ & $0.84 \pm 0.20$ & $0.98 \pm 0.06$ & $1.01 \pm 0.15$ & $1.10 \pm 0.17$ & $1.028 \pm 0.011$ & $1.7 \pm 1.1$ \\

A 133 & 0.2 & $3.476 \pm 0.021$ & $0.269 \pm 0.004$ & $0.86 \pm 0.13$ & $0.83 \pm 0.18$ & $0.57 \pm 0.07$ & $0.82 \pm 0.12$ & $0.8 \pm 0.4$ & $0.90 \pm 0.16$ & $1.25 \pm 0.19$ & $0.901 \pm 0.013$ & $2.1 \pm 1.4$ \\
         & 0.05 & $2.704 \pm 0.020$ & $0.227 \pm 0.004$ & $=$ & $=$ & $0.65 \pm 0.08$ & $0.89 \pm 0.04$ & $0.9 \pm 0.3$ & $1.1 \pm 0.5$ & $1.21 \pm 0.22$ & $1.29 \pm 0.03$ & $1.8 \pm 0.8$ \\

A 189 & 0.05 & $1.020 \pm 0.021$ & $0.178 \pm 0.016$ & $0.7 \pm 0.3$ & $0.27_{-0.21}^{+0.44}$ & $0.42 \pm 0.23$ & $1.4 \pm 0.7$ & $0.7 \pm 0.4$ & $0.08$ $(< 0.62)$ & $0.14$ $(< 1.44)$ & $0.88 \pm 0.13$ & $-$ \\

A 262 & 0.2 & $2.264 \pm 0.013$ & $0.192 \pm 0.003$ & $0.76 \pm 0.10$ & $0.93 \pm 0.22$ & $0.94 \pm 0.21$ & $0.93 \pm 0.11$ & $0.95 \pm 0.07$ & $0.85 \pm 0.13$ & $1.27 \pm 0.18$ & $0.726 \pm 0.012$ & $2.6 \pm 2.4$ \\
         & 0.05 & $1.957 \pm 0.015$ & $0.188 \pm 0.004$ & $=$ & $=$ & $0.59 \pm 0.08$ & $0.95 \pm 0.15$ & $1.01 \pm 0.09$ & $0.87 \pm 0.16$ & $1.2 \pm 0.8$ & $1.08 \pm 0.03$ & $0.9$ $(< 2.8)$ \\

A 496 & 0.2 & $3.754 \pm 0.012$ & $0.273 \pm 0.003$ & $1.03 \pm 0.12$ & $1.16 \pm 0.20$ & $0.86 \pm 0.20$ & $0.90 \pm 0.13$ & $0.96 \pm 0.05$ & $0.90 \pm 0.10$ & $1.0 \pm 0.4$ & $0.729 \pm 0.006$ & $1.3 \pm 0.3$ \\
         & 0.05 & $3.007 \pm 0.012$ & $0.229 \pm 0.003$ & $=$ & $=$ & $0.82 \pm 0.25$ & $0.94 \pm 0.13$ & $1.03 \pm 0.05$ & $0.97 \pm 0.11$ & $1.04 \pm 0.12$ & $0.947 \pm 0.010$ & $1.6 \pm 0.4$ \\

A 1795 & 0.2 & $5.15 \pm 0.04$ & $0.318 \pm 0.013$ & $1.7 \pm 0.4$ & $1.5 \pm 0.4$ & $0.63 \pm 0.15$ & $0.6 \pm 0.4$ & $1.1 \pm 0.6$ & $1.0 \pm 0.3$ & $1.4 \pm 0.4$ & $0.581 \pm 0.010$ & $2.1 \pm 0.6$ \\
            & 0.05 & $4.18 \pm 0.04$ & $0.298 \pm 0.013$ & $=$ & $=$ & $0.65 \pm 0.18$ & $0.8 \pm 0.5$ & $0.99 \pm 0.17$ & $1.4 \pm 0.4$ & $0.7 \pm 0.5$ & $0.683 \pm 0.018$ & $3.5 \pm 1.1$ \\

A 1991 & 0.2 & $2.28 \pm 0.03$ & $0.202 \pm 0.007$ & $0.9 \pm 0.3$ & $0.5 \pm 0.3$ & $0.66 \pm 0.15$ & $0.85 \pm 0.07$ & $1.18 \pm 0.20$ & $1.2 \pm 0.3$ & $1.6 \pm 0.4$ & $0.80 \pm 0.03$ & $4 \pm 3$ \\
            & 0.05 & $2.00 \pm 0.04$ & $0.205 \pm 0.009$ & $=$ & $=$ & $0.68 \pm 0.19$ & $0.81 \pm 0.09$ & $1.3 \pm 0.3$ & $1.2 \pm 0.4$ & $0.7 \pm 0.5$ & $1.14 \pm 0.07$ & $0.6$ $(< 2.9)$ \\

A 2029 & 0.2 & $7.48 \pm 0.04$ & $0.301 \pm 0.006$ & $1.70 \pm 0.25$ & $0.17$ $(< 0.59)$ & $0.5$ $(< 0.9)$ & $0.5 \pm 0.3$ & $0.8 \pm 0.3$ & $0.41 \pm 0.16$ & $0.96 \pm 0.20$ & $0.710 \pm 0.006$ & $1.58 \pm 0.22$ \\
            & 0.05 & $6.44 \pm 0.04$ & $0.292 \pm 0.009$ & $=$ & $=$ & $0.7 \pm 0.5$ & $0.7 \pm 0.3$ & $0.67 \pm 0.09$ & $0.39 \pm 0.20$ & $0.8$ $(< 1.5)$ & $0.885 \pm 0.012$ & $2.1 \pm 1.1$ \\

A 2052 & 0.2 & $2.931 \pm 0.012$ & $0.244 \pm 0.003$ & $0.84 \pm 0.09$ & $0.82 \pm 0.16$ & $0.57 \pm 0.06$ & $0.86 \pm 0.11$ & $0.8 \pm 0.4$ & $0.75 \pm 0.13$ & $1.15 \pm 0.16$ & $0.702 \pm 0.008$ & $2.6 \pm 0.7$ \\
            & 0.05 & $2.554 \pm 0.016$ & $0.234 \pm 0.004$ & $=$ & $=$ & $0.57 \pm 0.08$ & $0.92 \pm 0.22$ & $0.95 \pm 0.09$ & $0.91 \pm 0.18$ & $1.28 \pm 0.24$ & $0.917 \pm 0.017$ & $1.1 \pm 0.9$ \\

A 2199 & 0.2 & $4.111 \pm 0.011$ & $0.289 \pm 0.003$ & $1.12 \pm 0.14$ & $1.17 \pm 0.23$ & $0.76 \pm 0.05$ & $0.94 \pm 0.10$ & $0.92 \pm 0.05$ & $0.86 \pm 0.11$ & $1.27 \pm 0.13$ & $0.581 \pm 0.004$ & $2.3 \pm 0.9$ \\
            & 0.05 & $3.643 \pm 0.010$ & $0.272 \pm 0.003$ & $=$ & $=$ & $0.52 \pm 0.21$ & $0.99 \pm 0.14$ & $1.03 \pm 0.06$ & $1.02 \pm 0.14$ & $1.41 \pm 0.17$ & $0.774 \pm 0.005$ & $2.3 \pm 1.7$ \\

A 2597 & 0.2 & $3.421 \pm 0.016$ & $0.258 \pm 0.006$ & $1.30 \pm 0.20$ & $1.3 \pm 0.3$ & $0.54 \pm 0.11$ & $0.7 \pm 0.4$ & $1.06 \pm 0.13$ & $1.0 \pm 0.8$ & $1.1 \pm 0.3$ & $0.493 \pm 0.008$ & $2.1 \pm 1.9$ \\
            & 0.05 & $3.035 \pm 0.016$ & $0.242 \pm 0.007$ & $=$ & $=$ & $0.59 \pm 0.14$ & $0.6 \pm 0.4$ & $0.9 \pm 0.5$ & $1.4 \pm 1.1$ & $0.9 \pm 0.4$ & $0.564 \pm 0.011$ & $1.7 \pm 1.0$ \\

A 2626 & 0.2 & $3.06 \pm 0.03$ & $0.204 \pm 0.015$ & $0.9_{-0.3}^{+0.6}$ & $0.8 \pm 0.5$ & $0.42 \pm 0.16$ & $0.76 \pm 0.07$ & $0.99 \pm 0.18$ & $0.9 \pm 0.3$ & $0.9 \pm 0.4$ & $0.0687 \pm 0.019$ & $2.8 \pm 1.7$ \\
            & 0.05 & $2.77 \pm 0.05$ & $0.183 \pm 0.019$ & $=$ & $=$ & $0.8 \pm 0.3$ & $0.96 \pm 0.12$ & $0.55 \pm 0.20$ & $1.5 \pm 0.5$ & $0.23$ $(< 0.74)$ & $0.90 \pm 0.04$ & $-$ \\

A 3112 & 0.2 & $4.502 \pm 0.019$ & $0.289 \pm 0.005$ & $0.84 \pm 0.11$ & $0.42 \pm 0.20$ & $0.5 \pm 0.3$ & $0.64 \pm 0.14$ & $0.85 \pm 0.07$ & $0.71 \pm 0.13$ & $1.35 \pm 0.16$ & $0.814 \pm 0.006$ & $1.7 \pm 0.8$ \\
            & 0.05 & $3.696 \pm 0.019$ & $0.250 \pm 0.005$ & $=$ & $=$ & $0.5 \pm 0.3$ & $0.6 \pm 0.3$ & $0.76 \pm 0.06$ & $1.0 \pm 0.6$ & $1.22 \pm 0,18$ & $1.116 \pm 0.014$ & $1.7 \pm 1.2$ \\

A 3526 & 0.2 & $3.137 \pm 0.007$ & $0.2745 \pm 0.0012$ & $0.67 \pm 0.04$ & $0.74 \pm 0.10$ & $0.97 \pm 0.20$ & $0.99 \pm 0.07$ & $1.04 \pm 0.03$ & $0.9 \pm 0.3$ & $1.33 \pm 0.06$ & $1.104 \pm 0.005$ & $2.1 \pm 1.3$ \\
            & 0.05 & $2.554 \pm 0006$ & $0.2381 \pm 0.0010$ & $=$ & $=$ & $0.540 \pm 0.020$ & $0.99 \pm 0.08$ & $1.12 \pm 0.03$ & $1.08 \pm 0.04$ & $1.36 \pm 0.06$ & $1.675 \pm 0.011$ & $2.7 \pm 1.1$ \\

A 3581 & 0.05 & $1.637 \pm 0.008$ & $0.132 \pm 0.003$ & $0.77 \pm 0.08$ & $0.85 \pm 0.20$ & $0.75 \pm 0.18$ & $0.86 \pm 0.19$ & $1.19 \pm 0.08$ & $0.7 \pm 0.5$ & $1.37 \pm 0.19$ & $0.838 \pm 0.016$ & $0.0$ $-$ \\

A 4038 & 0.2 & $3.120 \pm 0.013$ & $0.225 \pm 0.005$ & $1.09 \pm 0.24$ & $0.3$ $(< 0.6)$ & $0.86 \pm 0.08$ & $0.86 \pm 0.18$ & $1.09 \pm 0.08$ & $0.9 \pm 0.6$ & $1.00 \pm 0.19$ & $0.547 \pm 0.007$ & $2.7 \pm 0.7$ \\
            & 0.05 & $3.117 \pm 0.021$ & $0.233 \pm 0.007$ & $=$ & $=$ & $0.5$ $(< 0.9)$ & $0.88 \pm 0.05$ & $0.98 \pm 0.11$ & $1.1 \pm 0.3$ & $1.3 \pm 0.3$ & $0.673 \pm 0.013$ & $3.5 \pm 1.1$ \\

A 4059 & 0.2 & $3.956 \pm 0.017$ & $0.294 \pm 0.004$ & $0.76 \pm 0.11$ & $0.91 \pm 0.21$ & $0.64 \pm 0.06$ & $0.85 \pm 0.15$ & $0.87 \pm 0.07$ & $0.78 \pm 0.14$ & $0.85 \pm 0.16$ & $0.756 \pm 0.008$ & $2.8 \pm 0.4$ \\
           & 0.05 & $3.275 \pm 0.018$ & $0.271 \pm 0.004$ & $=$ & $=$ & $0.67 \pm 0.09$ & $0.92 \pm 0.21$ & $0.97 \pm 0.09$ & $0.76 \pm 0.19$ & $0.98 \pm 0.22$ & $1.069 \pm 0.019$ & $3.4 \pm 0.8$ \\

AS 1101 & 0.2 & $2.503 \pm 0.007$ & $0.170 \pm 0.003$ & $0.75 \pm 0.10$ & $0.69 \pm 0.14$ & $0.7 \pm 0.3$ & $0.76 \pm 0.17$ & $0.81 \pm 0.07$ & $0.76 \pm 0.13$ & $0.95 \pm 0.16$ & $0.503 \pm 0.005$ & $2.1$ $(< 4.2)$ \\
               & 0.05 & $2.359 \pm 0.011$ & $0.163 \pm 0.005$ & $=$ & $=$ & $0.63 \pm 0.09$ & $0.79 \pm 0.19$ & $0.88 \pm 0.08$ & $0.91 \pm 0.17$ & $1.08 \pm 0.21$ & $0.606 \pm 0.010$ & $2.5 \pm 1.0$ \\

AWM 7 & 0.2 & $3.753 \pm 0.010$ & $0.240 \pm 0.004$ & $1.20 \pm 0.18$ & $0.4 \pm 0.3$ & $0.82 \pm 0.15$ & $0.91 \pm 0.09$ & $0.96 \pm 0.04$ & $0.9 \pm 0.3$ & $1.32 \pm 0.09$ & $0.781 \pm 0.004$ & $2.4 \pm 1.2$ \\
            & 0.05 & $3.528 \pm 0.014$ & $0.238 \pm 0.004$ & $=$ & $=$ & $0.52 \pm 0.21$ & $0.97 \pm 0.11$ & $1.14 \pm 0.05$ & $1.09 \pm 0.10$ & $1.3 \pm 0.4$ & $1.171 \pm 0.011$ & $2.4 \pm 1.0$ \\

EXO 0422 & 0.2 & $2.904 \pm 0.020$ & $0.182 \pm 0.011$ & $1.1 \pm 0.3$ & $0.9 \pm 0.4$ & $0.64 \pm 0.12$ & $0.91 \pm 0.05$ & $0.95 \pm 0.13$ & $0.87 \pm 0.23$ & $1.30 \pm 0.28$ & $0.633 \pm 0.013$ & $1.9 \pm 1.2$ \\
                 & 0.05 & $2.655 \pm 0.023$ & $0.162 \pm 0.014$ & $=$ & $=$ & $0.63 \pm 0.14$ & $0.97 \pm 0.07$ & $0.96 \pm 0.13$ & $0.9 \pm 0.3$ & $1.4 \pm 0.3$ & $0.817 \pm 0.023$ & $-$ \\

Fornax & 0.05 & $1.321 \pm 0.007$ & $0.15 \pm 0.004$ & $0.64 \pm 0.10$ & $0.56 \pm 0.19$ & $0.82 \pm 0.06$ & $1.0 \pm 0.3$ & $1.2 \pm 0.6$ & $1.0 \pm 0.6$ & $0.9 \pm 0.3$ & $0.856 \pm 0.021$ & $-$ \\

HCG 62 & 0.05 & $0.958 \pm 0.007$ & $0.128 \pm 0.004$ & $0.77 \pm 0.11$ & $1.14 \pm 0.23$ & $0.79 \pm 0.09$ & $1.0 \pm 0.3$ & $1.3 \pm 0.6$ & $1.4 \pm 0.4$ & $0.8$ $(< 1.6)$ & $0.70 \pm 0.04$ & $-$ \\

Hydra A & 0.2 & $3.450 \pm 0.010$ & $0.269 \pm 0.003$ & $1.3 \pm 0.3$ & $0.53_{-0.06}^{+0.26}$ & $0.42 \pm 0.09$ & $0.72 \pm 0.24$ & $0.9 \pm 0.4$ & $0.75 \pm 0.18$ & $1.00 \pm 0.22$ & $0.488 \pm 0.005$ & $2.1 \pm 0.5$ \\
             & 0.05 & $3.303 \pm 0.015$ & $0.257 \pm 0.005$ & $=$ & $=$ & $0.33 \pm 0.10$ & $0.69 \pm 0.29$ & $1.0 \pm 0.3$ & $0.9$ $(< 1.6)$ & $0.5 \pm 0.3$ & $0.601 \pm 0.010$ & $3.1 \pm 0.8$ \\

M 49 & 0.05 & $1.148 \pm 0.004$ & $0.269 \pm 0.003$ & $0.64 \pm 0.10$ & $0.78 \pm 0.20$ & $0.83 \pm 0.23$ & $1.04 \pm 0.13$ & $1.18 \pm 0.09$ & $1.01 \pm 0.16$ & $1.0 \pm 0.3$ & $0.928 \pm 0.022$ & $-$ \\

M 60 & 0.05 & $0.923 \pm 0.003$ & $0.042 \pm 0.004$ & $0.66 \pm 0.07$ & $0.66 \pm 0.13$ & $1.2 \pm 0.4$ & $1.1 \pm 0.3$ & $0.8 \pm 0.3$ & $1.0 \pm 0.4$ & $0.4$ $(< 1.5)$ & $0.478 \pm 0.011$ & $-$ \\

M 84 & 0.05 & $0.99 \pm 0.03$ & $0.36 \pm 0.06$ & $1.13 \pm 0.24$ & $0.22 \pm 0.18$ & $2.7 \pm 2.0$ & $1.4 \pm 0.6$ & $2.6 \pm 2.2$ & $1.3$ $(< 2.7)$ & $0.0$ $(< 1.8)$ & $0.37 \pm 0.03$ & $-$ \\

M 86 & 0.05 & $0.967 \pm 0.006$ & $0.120 \pm 0.004$ & $1.23 \pm 0.24$ & $0.32_{-0.09}^{+0.47}$ & $1.22 \pm 0.08$ & $0.96 \pm 0.21$ & $1.1 \pm 0.4$ & $1.0 \pm 0.3$ & $1.8$ $(< 3.6)$ & $0.538 \pm 0.019$ & $-$ \\

M 87 & 0.05 & $2.0517 \pm 0.0018$ & $0.1572 \pm 0.0005$ & $1.06 \pm 0.17$ & $0.61 \pm 0.12$ & $0.707 \pm 0.014$ & $1.13 \pm 0.05$ & $1.30 \pm 0.16$ & $1.21 \pm 0.20$ & $1.55 \pm 0.16$ & $0.7546 \pm 0.0014$ & $2.6 \pm 0.4$ \\

M 89 & 0.05 & $0.646 \pm 0.018$ & $0.10*$ & $2.0 \pm 0.6$ & $1.1 \pm 0.7$ & $-$ & $1.4_{-0.7}^{+5.3}$ & $-$ & $-$ & $-$ & $0.30 \pm 0.07$ & $-$ \\

MKW 3s & 0.2 & $3.637 \pm 0.013$ & $0.254 \pm 0.004$ & $0.84 \pm 0.23$ & $0.72 \pm 0.22$ & $0.9 \pm 0.3$ & $0.84 \pm 0.18$ & $0.89 \pm 0.07$ & $0.84 \pm 0.16$ & $1.29 \pm 0.18$ & $0.607 \pm 0.007$ & $2.0 \pm 0.4$ \\
              & 0.05 & $3.379 \pm 0.018$ & $0.232 \pm 0.006$ & $=$ & $=$ & $0.77 \pm 0.10$ & $0.81 \pm 0.25$ & $0.97 \pm 0.09$ & $0.75 \pm 0.19$ & $1.01 \pm 0.23$ & $0.863 \pm 0.013$ & $0.9 \pm 0.5$ \\

MKW 4 & 0.2 & $1.909 \pm 0.011$ & $0.121 \pm 0.004$ & $0.66 \pm 0.12$ & $0.87 \pm 0.26$ & $0.69 \pm 0.07$ & $0.97 \pm 0.14$ & $1.0 \pm 0.3$ & $0.92 \pm 0.14$ & $0.85 \pm 0.20$ & $1.008 \pm 0.013$ & $-$ \\
            & 0.05 & $1.756 \pm 0.012$ & $0.116 \pm 0.004$ & $=$ & $=$ & $0.9 \pm 0.4$ & $1.04 \pm 0.16$ & $1.4 \pm 0.4$ & $1.22 \pm 0.17$ & $0.89 \pm 0.23$ & $1.64 \pm 0.03$ & $0.0$ $(< 1.4)$ \\

NGC 507 & 0.05 & $1.257 \pm 0.008$ & $0.131 \pm 0.005$ & $0.69_{-0.18}^{+0.39}$ & $0.14_{-0.06}^{0.60}$ & $0.73 \pm 0.08$ & $0.9 \pm 0.3$ & $1.3 \pm 0.5$ & $1.46 \pm 0.25$ & $1.0 \pm 0.4$ & $0.89 \pm 0.03$ & $0.0$ $(< 3.0)$ \\

NGC 1316 & 0.05 & $0.768 \pm 0.012$ & $0.235 \pm 0.016$ & $1.2 \pm 0.4$ & $1.0 \pm 0.4$ & $1.0 \pm 0.9$ & $0.66 \pm 0.12$ & $0.9 \pm 0.6$ & $0.7$ $(< 2.3)$ & $-$ & $0.34 \pm 0.03$ & $-$ \\

NGC 1404 & 0.05 & $0.64 \pm 0.03$ & $0.359 \pm 0.04$ & $0.89 \pm 0.22$ & $0.60 \pm 0.23$ & $-$ & $0.8 \pm 0.4$ & $-$ & $-$ & $-$ & $0.61 \pm 0.10$ & $-$ \\

NGC 1550 & 0.05 & $1.328 \pm 0.004$ & $0.088 \pm 0.003$ & $0.95 \pm 0.11$ & $0.80 \pm 0.20$ & $0.68 \pm 0.04$ & $0.95 \pm 0.19$ & $1.09 \pm 0.22$ & $1.1 \pm 0.6$ & $1.17 \pm 0.20$ & $0.607 \pm 0.010$ & $-$ \\

NGC 3411 & 0.05 & $0.933 \pm 0.006$ & $(<0.018)$ & $0.71 \pm 0.23$ & $0.7 \pm 0.5$ & $1.3 \pm 0.5$ & $1.0 \pm 0.3$ & $1.13 \pm 0.23$ & $0.8 \pm 0.6$ & $0.0$ $(< 1.7)$ & $0.59 \pm 0.03$ & $-$ \\

NGC 4261 & 0.05 & $0.940 \pm 0.021$ & $0.32 \pm 0.04$ & $0.95\pm 0.25$ & $0.9 \pm 0.3$ & $1.2 \pm 0.6$ & $1.2 \pm 0.5$ & $1.3 \pm 1.0$ & $-$ & $1.7$ $(< 4.9)$ & $0.35 \pm 0.03$ & $-$ \\

NGC 4325 & 0.05 & $0.931 \pm 0.012$ & $0.071 \pm 0.010$ & $0.65 \pm 0.18$ & $1.1 \pm 0.4$ & $0.50 \pm 0.16$ & $0.80 \pm 0.11$ & $1.2 \pm 0.4$ & $1.7 \pm 1.0$ & $1.9$ $(< 4.1)$ & $0.64 \pm 0.06$ & $-$ \\

NGC 4636 & 0.05 & $0.7498 \pm 0.0025$ & $0.103 \pm 0.004$ & $0.87 \pm 0.07$ & $0.71 \pm 0.10$ & $1.19 \pm 0.10$ & $1.1 \pm 0.3$ & $1.23 \pm 0.15$ & $1.6_{-0.6}^{+1.6}$ & $-$ & $0.82 \pm 0.06$ & $-$ \\

NGC 5044 & 0.05 & $0.9741 \pm 0.0020$ & $0.0811 \pm 0.0011$ & $0.85 \pm 0.07$ & $0.74 \pm 0.14$ & $1.00 \pm 0.03$ & $0.93 \pm 0.14$ & $1.3 \pm 0.3$ & $1.4 \pm 0.5$ & $1.5 \pm 0.3$ & $0.668 \pm 0.010$ & $-$ \\

NGC 5813 & 0.05 & $0.7275 \pm 0.0016$ & $0.092 \pm 0.003$ & $0.76 \pm 0.07$ & $0.42 \pm 0.09$ & $1.45 \pm 0.09$ & $1.1 \pm 0.3$ & $1.2 \pm 0.5$ & $1.0 \pm 0.6$ & $0.0$ $(< 0.6)$ & $0.715 \pm 0.020$ & $-$ \\

NGC 5846 & 0.05 & $0.7585 \pm 0.0023$ & $0.128 \pm 0.004$ & $0.99 \pm 0.11$ & $0.90 \pm 0.15$ & $0.89 \pm 0.14$ & $1.3 \pm 0.3$ & $1.22 \pm 0.13$ & $1.3 \pm 0.4$ & $0.6$ $(< 1.4)$ & $0.659 \pm 0.017$ & $-$ \\

Perseus & 0.2 & $4.865 \pm 0.004$ & $0.3177 \pm 0.0006$ & $1.24 \pm 0.18$ & $1.05 \pm 0.16$ & $0.74 \pm 0.09$ & $0.83 \pm 0.03$ & $0.87 \pm 0.09$ & $0.9 \pm 0.3$ & $0.97 \pm 0.18$ & $0.6934 \pm 0.0012$ & $1.8 \pm 0.5$ \\
              & 0.05 & $3.905 \pm 0.003$ & $0.2752 \pm 0.0007$ & $=$ & $=$ & $0.45 \pm 0.15$ & $0.81 \pm 0.04$ & $0.90 \pm 0.11$ & $0.9 \pm 0.3$ & $1.04 \pm 0.18$ & $0.7543 \pm 0.0015$ & $1.9 \pm 0.6$ \\
\hline
\end{longtable}
\end{landscape}
\end{longtab}

\end{document}